\documentclass[review]{elsarticle}
\usepackage{lineno}
\usepackage{geometry}
\usepackage{booktabs}
\usepackage{threeparttable}
\usepackage{amsmath}
\usepackage{amsthm}
\usepackage{mathrsfs}
\usepackage{amssymb}
\usepackage{graphicx}
\usepackage{multirow}
\usepackage{subfig}
\usepackage{bm}
\usepackage{algorithm}
\usepackage{algpseudocode}
\usepackage{graphics}
\usepackage[colorlinks,linkcolor=red, anchorcolor=blue,citecolor=green]{hyperref}

\modulolinenumbers[5]
\journal{Journal of \LaTeX\ Templates}
\graphicspath{{./}}

\biboptions{authoryear}
\DeclareMathOperator*{\argmax}{arg\,max}
\begin{document}
\begin{frontmatter}
\title{Investigation of Pedestrian Dynamics in Circle Antipode Experiments} %%%%%%%%%%%%%%%%%%%%%%%%%%%%%%%%%%%%%%%%%
% \author{Yao Xiao, Rui Jiang, Ziyou Gao\corref{mycorrespondingauthor}, Yunchao Qu, Xingang Li}
% \address{Institute of Transportation System Science and Engineering, Beijing Jiaotong University, Beijing}

\author[]{Yao Xiao}
\author[]{Ziyou Gao\corref{mycorrespondingauthor}}
\author[]{Rui Jiang}
\author[]{Xingang Li}
\author[]{Yunchao Qu}
\address{School of traffic and transportation, Beijing Jiaotong University, China}
\cortext[mycorrespondingauthor]{E-mail address: zygao@bjtu.edu.cn} 

% \fntext[myfootnote]{Since 1880.}
% %% or include affiliations in footnotes:
% \author[mymainaddress,mysecondaryaddress]{Elsevier Inc}
%\author[1222]{\corref{mycorrespondingauthor}}
% \address[mysecondaryaddress]{360 Park Avenue South, New York}

\begin{abstract}
To explore the pedestrian motion navigation and conflict reaction mechanisms in practice, we organized a series of circle antipode experiments. In the experiments, pedestrians are uniformly initialized on the circle and required to leave for their antipodal positions simultaneously. On the one hand, a conflicting area is naturally formulated in the center region due to the converged shortest routes, so the practical conflict avoidance behaviors can be fully explored. On the other hand, the symmetric experimental conditions of pedestrians, e.g., symmetric starting points, symmetric destination points, and symmetric surroundings, lay the foundation for further quantitative comparisons among participants. The pedestrian trajectories in the experiments are recognized and rotated, and several aspects, e.g., the trajectory space distribution, route length, travel time, velocity distribution, and time-series, are investigated. It is found that: (1) Pedestrians prefer the right-hand side during the experiments; (2) The route length is as the law of log-normal distribution, the route potential obeys the exponential distribution, and the travel time is normally distributed as well as the speed; (3) Taking the short routes unexpectedly cost pedestrians plenty of travel time, while detour seems to be a time-saving decision. 

What's more, the series of experiments can be regarded as a basis of the model evaluation benefit from the serious conflicts and the symmetric conditions. The evaluation framework contains four distribution indexes and two time series indexes in space and time dimensions, and they are respectively graded according to the K-S test and the DTW method. A traditional social force model and a Voronoi diagram based modification are introduced to test the evaluation framework. The evaluation results show that the framework is beneficial to evaluate pedestrian models and even reflects the minor differences between the models. 

%为了揭示行人在实际条件甚至较复杂条件下的冲突避让和拥挤处理机制，我们组织了一系列的重复圆形对角走行实验。之于我们的问题，该实验具有两方面的显著特点。一方面，人群的最短路径汇聚于实验的圆形中心点，在实验过程之中人群会在实验区域的中部形成一个拥挤区域，而我们可以充分的观察该拥挤条件下行人的实际应对行为。另一方面，各个行人的实验条件高度相同，包括均匀对称分布的起点位置，终点位置和实验任务。而这一特点可以使得我们充分利用所有人的行动轨迹，并且为更进一步的量化分析提供了坚实基础。考虑这些特点，我们定义并研究分析了实验中行人行为的诸多特征，轨迹空间分布，路径长度，旅程时间，速度分布以及时间序列的相关参数。其中我们发现了，一、行人的右行倾向性；二、路径长度的对数正态分布和旅程时间的正态分布；三、较短路径的平均行程时间更长，而较长路径的平均行程时间更短。除此之外，该实验还可以用来作为行人流模型评价的基础。本文评价框架包括了空间和时间相关的六个参数。而根据参数的不同特性，我们分别提出了基于K-S test 和 DTW distance 的方法评价各个参数。同时，本文还分别基于传统的社会力模型以及一个基于Voronoi图的改进社会模型中应用了该评价体系。结果发现，该评价体系可以较好的评价模型，并且体现模型之间的差异。
% 服从指数分布  it obeys the exponential distribution

\end{abstract}

\begin{keyword}
Pedestrian Dynamics, Circle Antipode Experiment, Pedestrian Trajectories, Model Evaluation
%\texttt{elsarticle.cls}\sep \LaTeX\sep Elsevier \sep template
%\MSC[2010] 00-01\sep 99-00
\end{keyword}
\end{frontmatter}

\section{Introduction}\label{section 1}

The research of pedestrian dynamics is attracting more attention in recent decades due to the increasing frequency of large-scale events around the world. The researches could be applied for the optimization of public facilities as well as the organization of pedestrian crowds. The investigation of experimental data is one of the most effective methods to explore the pedestrian dynamics \citep{Helbing2005, Seyfried2009, Haghani2018}. Typical data collection approaches include the controlled experiment with humans, animal experiment, virtual reality and hypothetical choice experiment, evacuation drill experiment, natural disaster analysis, natural environment analysis, and post-disaster interview \citep{Haghani2018}.

Among the data collection approaches, the controlled experiment with humans is a widely used method in the study of pedestrian dynamics, and numbers of scenarios (e.g., uni-directional flow, bottleneck, multi-directional flow, etc.) have been applied in the investigation. The single line movement shall be the most fundamental scenario in the uni-directional flow scenarios. \cite{Seyfried2005} investigated the single line experiment and found the linear relation between the velocity and the inverse of the density. The kindred experiment performed by \cite{Jelic2012a} drew similar results, and a further investigation on the stepping behaviors \citep{Jelic2012} showed that the step length was positively correlated with the pedestrian velocity, and the available space in front was negatively correlated with the variations of the step duration. In the uni-directional flow experiments, plenty of experimental results between macroscopic parameters, i.e., density, velocity and flow, have been accumulated \citep{Hankin1958,Older1968,Mori1987,Weidmann1993,Helbing2007}, and \cite{Zhang2011} performed a comparison between four different measurement methods and found that the Voronoi method could resolve the fine structure of the fundamental diagram.

The bottleneck experiment considers a different scenario which explores the pedestrian behaviors and reactions in crowded situations. The arching phenomenon \citep{Predtechenskii1978} and zipper effect \citep{Hoogendoorn2005,Seyfried2009} are typical self-organized phenomena in the type of experiment. The calculated capacities in the experiments \citep{Predtechenskii1978,Kretz2006,Seyfried2009} vary from $1.3 s^{-1}$ to $1.9 s^{-1}$, and researchers have deeply investigated the mechanisms of the behavior characteristics. \cite{Nicolas2017} presented a further investigation on heterogeneous behaviors, namely selfish and polite attitude. It was found that the growing ratio of selfish pedestrians led to the rise and the disorder of flow rate. \cite{Kruechten2017} investigated the behaviors of social groups and concluded that the increasing size of social groups could have a positive influence on evacuation.

The multi-directional flow experiment can achieve more interesting results by applying the complicated scenarios with more conflict avoidances and congestion reactions. To our knowledge, the experiments mainly include the bi-directional flow in a corridor, T-junction and crossing flow. In the bi-directional flow of a corridor, the lane formation \citep{Daamen2003,Helbing2005,Zhang2012,Moussaid2012} is the most well-known self-organized phenomenon, and the formulated unidirectional lane is generally recognized as a promotion for an efficient motion. Distinguished differences between the bi- and uni-directional flow experiments can be found in the fundamental diagrams \citep{Older1968,Navin1969,Polus1983,Tanaboriboon1986,Weidmann1993,Lam2002,Guo2012a,Zhang2012,Cao2017}. The T-junction and crossing flow experiments with different intersection angles \citep{Daamen2003,Helbing2005,Asano2010,Cao2017} are also conducted, and the phenomenon named stripe formation \citep{Ando1988,Helbing2005} has been observed. Other frequently-used scenarios for controlled experiments include circular area experiments\citep{Dyer2008,Dyer2009,Faria2009,Faria2010} and route choice experiments \citep{Guo2012, Haghani2017, Wagoum2017}.

% The point is that the motivation of the experiment is not appropriately described. 
% To fix it, we have to modify the construction of the section. 

\begin{figure}[!ht]
\centering\includegraphics[width=16 cm]{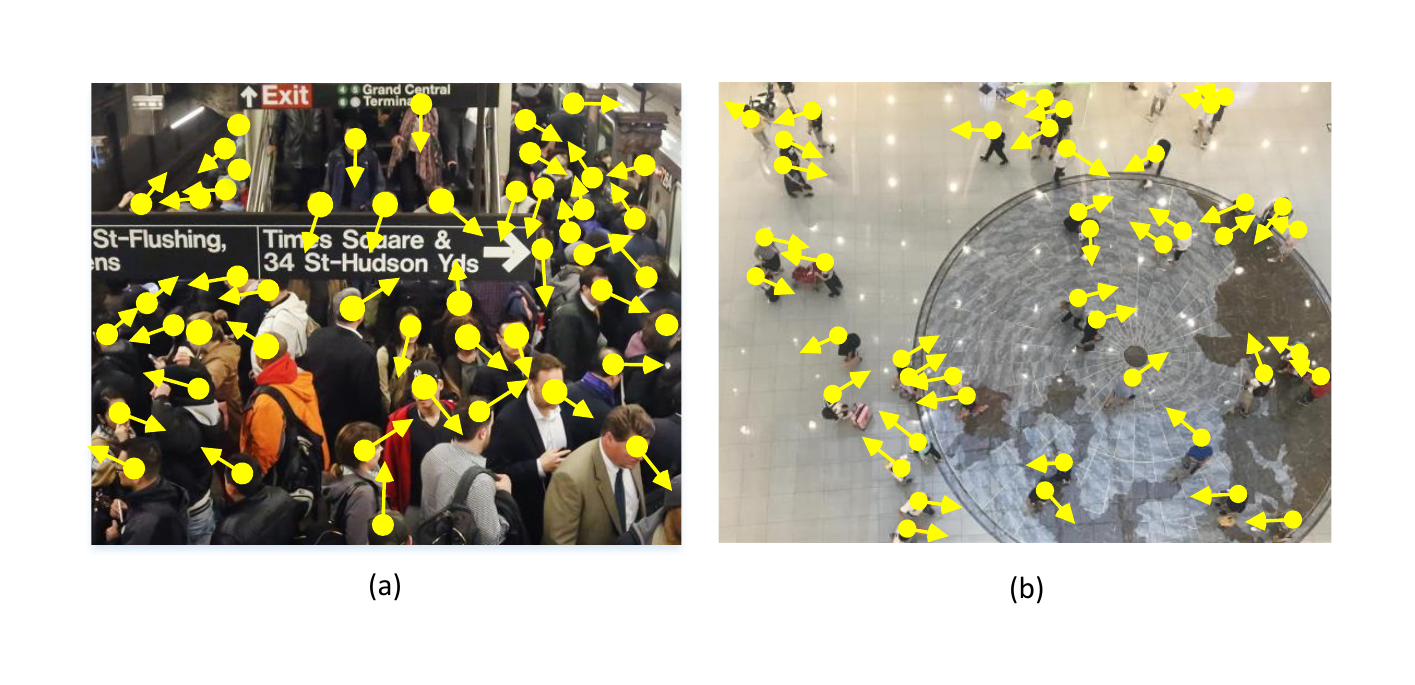}
\caption{Pedestrian crowd in practice. The yellow arrow represents the walking direction of a pedestrian. (a) pedestrians in a subway station. (b) pedestrians in a shopping mall.} \label{figcrowd}
\end{figure}
 
Nevertheless, due to the variances of trip destinations and individual preferences, the walking directions of pedestrians in practice usually spread over almost all angles and there is no dominant direction on the whole (see Fig. \ref{figcrowd}). In this situation, complex and serious conflicts frequently emerge among pedestrians. How to deal with the complex and serious conflicts shall be a core and challenging part in the investigation of pedestrian dynamics. Although some conflicts can be found in current experiments such as the bi- and multi-directional flow experiments, following behaviors are adopted by most pedestrians to avoid conflicts. As a result, the mechanisms of route navigation and conflict avoidance are not fully revealed in the current experiments. Another problem in most of the pedestrian experiments is the limitation on the quantitative investigations. Many performance indicators of participants, such as the route length and travel time, are difficult to be quantitatively compared and analyzed due to the variances of experiment conditions (e.g., initial positions, destination positions). Indeed, the method of repeated experiments is still possible for the achievement of general behaviors \citep{Moussaid2009}. However, for most large-scale experiments of human beings, the repeated method is costly and usually inefficient since the performances of participants are still hard to compare. Therefore, designing an experiment with symmetric experimental conditions for participants is of great significance for a further quantitative investigation. In the case, the circle antipode experiment draws our attention.

The circle antipode experiment \citep{VandenBerg2008,Ondrej2010,Golas2014} has been applied as a simulation scene, but very limited attention is paid to the performance of pedestrians in reality. In the circle antipode experiments, pedestrians are uniformly initialized on a circle, and they are required to reach the antipodal positions. The first significant feature of the experiment is that the shortest routes intersect at the center of the circle, and a crowded area is generally formulated in the center zone. In the case, a pedestrian has to deal with the conflicts with other pedestrians and pass through the crowded area in front. The ability to deal with the situations of conflicts and congestions shall be a fundamental and core problem in the pedestrian research, and here the practical handling ability of conflicts and collisions could be greatly revealed. Another significant but rarely mentioned feature is that each pedestrian is owning a symmetric initial position and a symmetric destination position in the experiment. In other words, except for the heterogeneity of pedestrians themselves, each pedestrian faces a symmetric experimental condition, namely a symmetric initial position, symmetric destination position etc. Based on it, the results of pedestrian trajectories could be used to the maximum possible extent, and more quantitative investigations are possible. Finally, with the development of the video recognition technology and other trajectory recognition technologies in recent years \citep{Boltes2010,Boltes2013,Seer2014,Corbetta2014}, obtaining the precise trajectories of pedestrians even in crowded situations is not so difficult as before. The achievement of the precise pedestrian trajectories provides more detailed characteristics of microscopic pedestrian behaviors. In summary, the access to the trajectories of the circle antipode experiments provides more possibilities for quantitatively investigating the performance of pedestrians in a challenging conflicting situation.

What's more, the circle antipode experiments show the potential to be an evaluation basis for pedestrian simulation models. An appropriate and quantitative evaluation is critical for the calibration, validation, optimization, and comparison of the pedestrian simulation models. Popular evaluation methods for the pedestrian simulation model mainly include the applications of self-organized phenomena, fundamental diagrams, and trajectories. \cite{Helbing1995} found and analyzed the reproduction of lane formation in the social force model. The performance of the generalized centrifugal force model was discussed on the basis of the fundamental diagram \citep{Chraibi2010}. Individual trajectories and collective pattern were applied for the validation of pedestrian models \citep{Antonini2006, Robin2009}. Also, a combination of the three types of methods could be adopted in the pedestrian models for evaluation \citep{Moussaid2012, Xiao2016}. Other evaluation methods include the model applicability for route choices and computational burden \citep{Duives2013}.

In general, lots of models can meet the requirements of the above evaluation methods. Even so, these models are probably unable to reproduce complete realistic pedestrian behaviors, especially in crowded and complicated situations. A key reason is the lack of tests on the collision avoidance ability which shall be quite significant in practice. In the circle antipode experiments, the challenging crowded situation makes the experiment an ideal scenario for exploring the conflict avoidance behaviors, and the characteristic of symmetric experimental condition provides more room for the quantitative evaluation. Therefore, the circle antipode experiment can be applied as an evaluation basis for the pedestrian models.

Here, a series of circle antipode experiments were performed with different numbers of pedestrians and different radii of circles, and the practical pedestrian trajectories on the ground were precisely extracted through video technologies. Based on the symmetric and uniform distribution of pedestrians, the original trajectories are rotated to make the starting points and the destination points overlapped, respectively. As a result, the normalized rotations allow more quantitative investigations of pedestrian dynamics, and the performances of pedestrians including trajectory spatial distribution, route length, travel time, velocity, and time-dependent features are analyzed. Considering the serious conflicting situations and the symmetric experimental conditions in the circle antipode experiments, an evaluation framework is proposed for the calibration, validation, optimization, and comparison of pedestrian simulation models.

The rest of paper is organized as follows. Section \ref{section 2} describes the detailed settings of the circle antipode experiments. In Section \ref{section 3}, the experiment results especially the pedestrian trajectories are analyzed. In Section \ref{section 4}, an evaluation framework for the pedestrian trajectories in the circle antipode experiment is developed for the model evaluation. With the evaluation framework, a traditional social force model and a Voronoi based social force model are proposed and tested. Finally, major conclusions and problems are drawn, and future researches are pointed out in section \ref{section 5}.

\section{Circle antipode experiments}\label{section 2}

\begin{figure}[!ht]
\centering 
%\hspace{-0ex} \vspace{-1ex}
\subfloat[]{ \label{figcirclegamea}\includegraphics[width=0.3\textwidth]{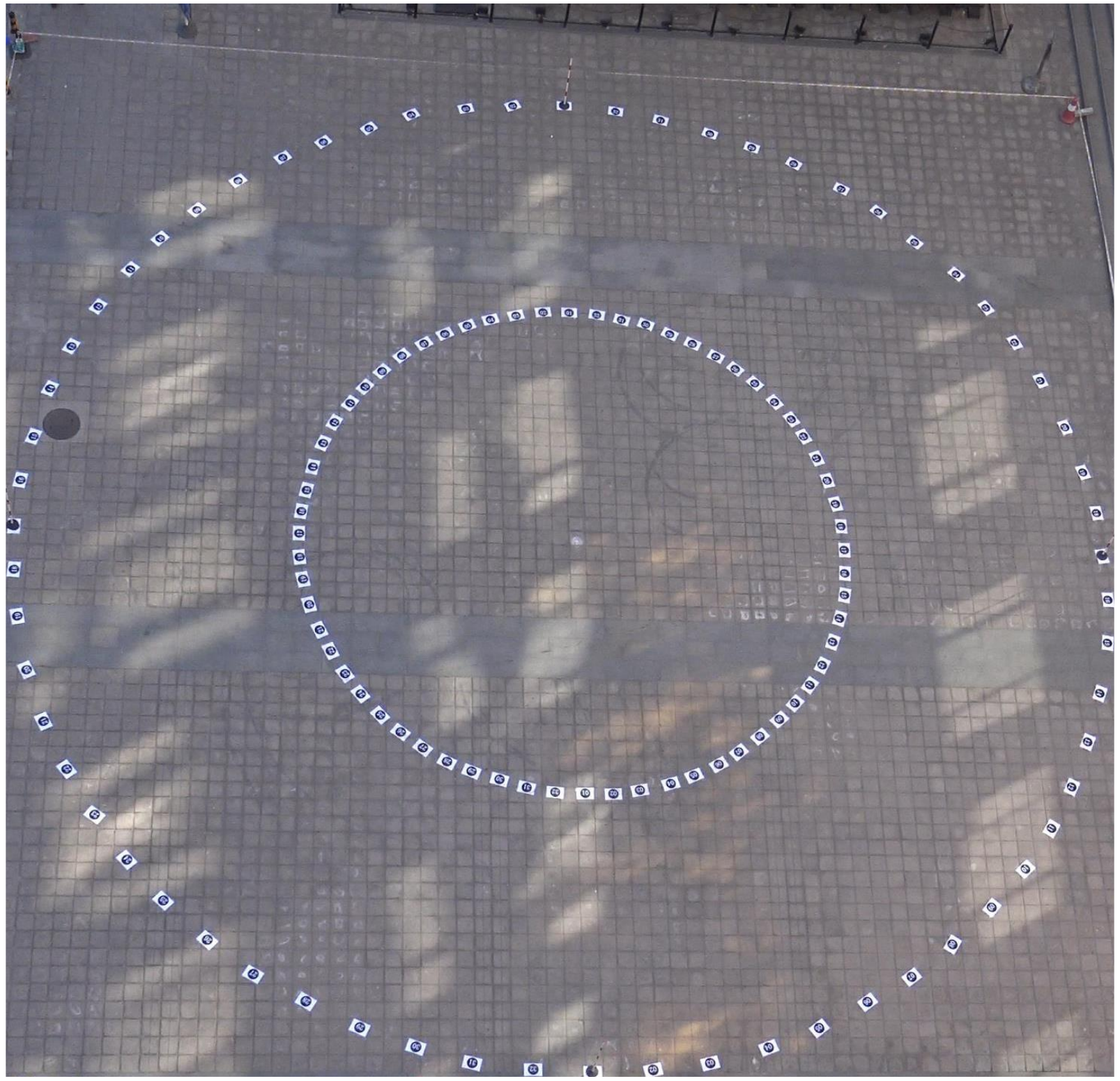}}
%\hspace{-6ex}
\subfloat[]{ \label{figcirclegameb}\includegraphics[width=0.3\textwidth]{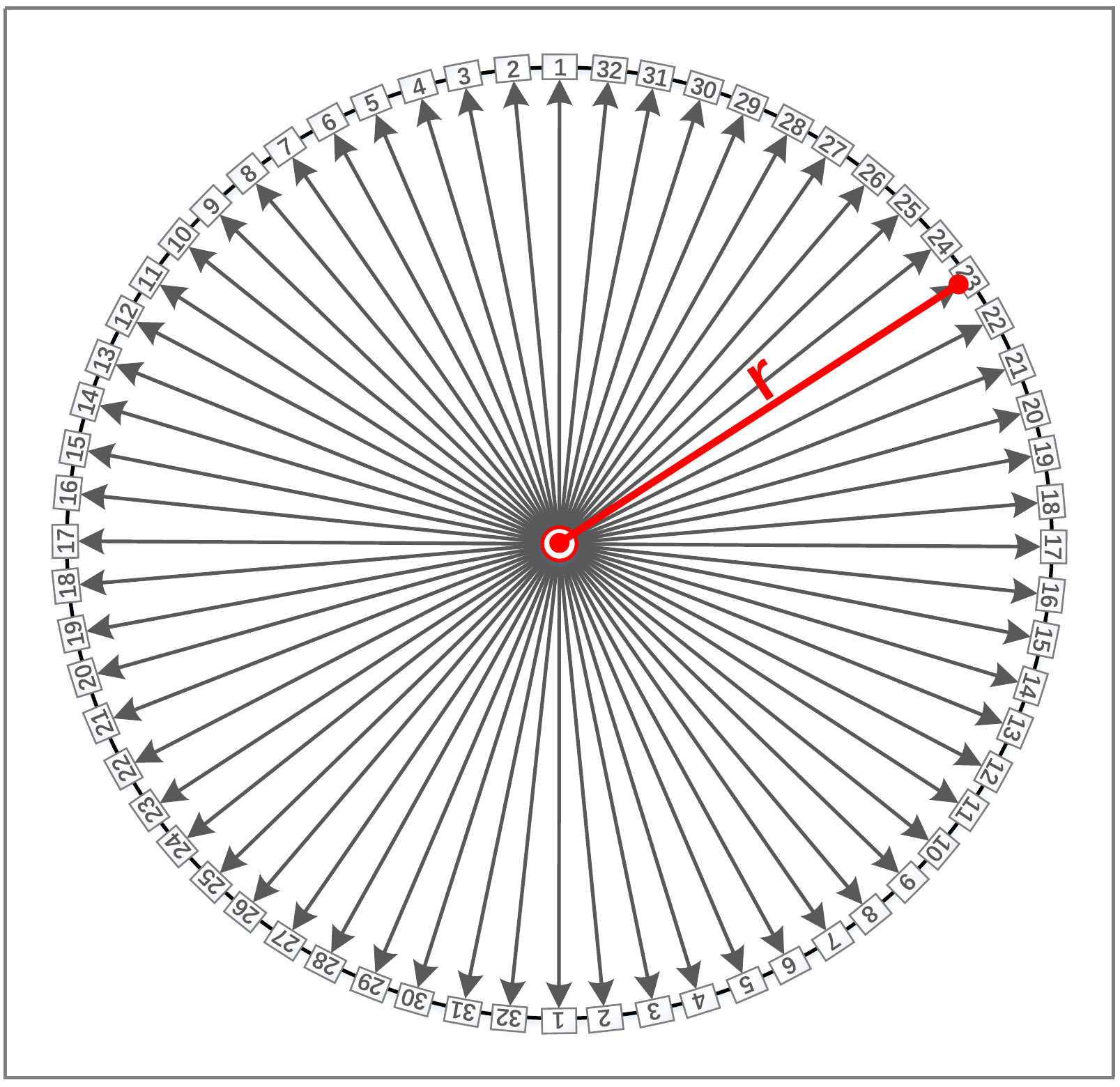}}
\subfloat[]{ \label{figcirclegamec}\includegraphics[width=0.33\textwidth]{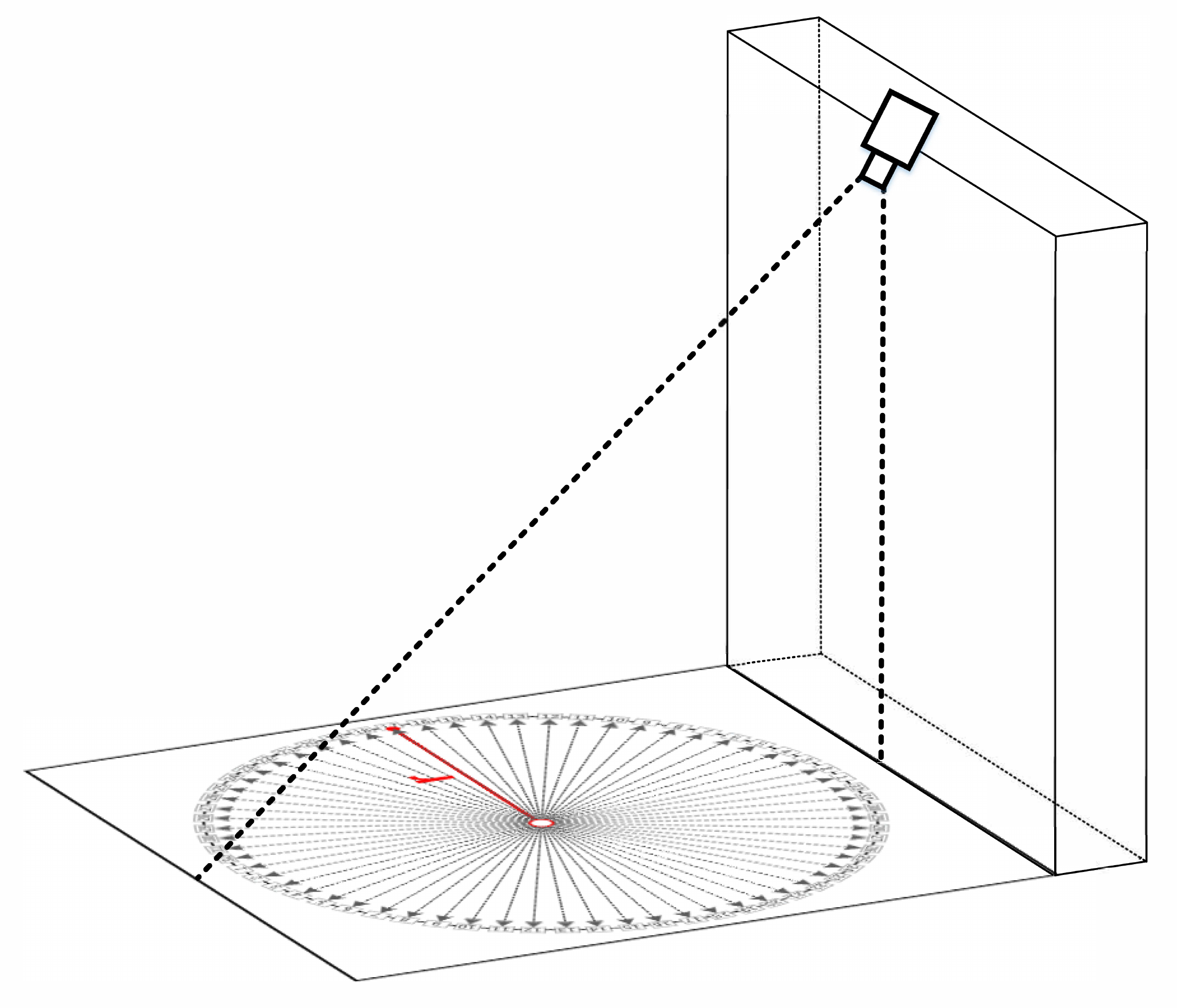}}
\caption{Circle antipode experiment. (a) Snapshot of the experimental square. (b) Sketch map of circle antipode experiment. (c) Illustration of camera position.} \label{figcirclegame} 
\end{figure}

The circle antipode experiment was conducted on a public square in front of a teaching building in Beijing Jiaotong University, China. On the ground of the public square, two circles of radii 5m and 10m were respectively plotted, and four sets of surface marks (totaling 128) from 1 to 32 were uniformly pasted on the two circles (see Fig. \ref{figcirclegamea}). The surface marks were used as the starting points and the destination points for the participants, hence they were specifically designed to guide the pedestrians to the correct position without hesitation or deviation. Considering the set up of the surface marks and the uniform distribution of experimental participants, $64=2^6$ was selected as the maximal number of participants in the experiment. A sketch map for the experiment setting can be found in Fig. \ref{figcirclegameb}. A surface mark and its antipodal mark shared a common number. Besides, a square area, labeled with the cordon, was used to limit the feasible movement area for experimental pedestrians and prevent unrelated pedestrians from entering.

Up to 64 participants (31 females, 33 males) were involved in the experiment, and they were students from the university and aged 18 to 28. For the sake of video recognition, participants were asked to wear clothes with the color of black, white or gray. Besides, each participant was required to wear a specific hat with featured color according to his/her height. For instance, the blue, red and yellow hats corresponded to the participant height from 155 cm to 165 cm, from 165 cm to 175 cm, and from 175 cm to 185 cm, respectively. Here, the numbers of these three groups of participants were 21, 29, and 14, respectively. The 64 participants were randomly divided into two groups (A and B) with the same size, and the participants in each group were located on a semicircle. In group A or B, each participant was randomly assigned with a number from 1 to 32, and he/she would stand on the surface mark with the assigned number at the initial phase. Once receiving the start order, the participants would leave for their antipodal positions on the circle as quickly as possible. Due to the potential jam and security problem in the center region, the participants were warned to take care of the security issues during the experiments.

\begin{table}[!ht]
\centering
\caption{Experiment arrangement.} \label{tab1}
\begin{tabular}{cccccccc}
\toprule
\multicolumn{4}{c}{5m experiment} & \multicolumn{4}{c}{10m experiment} \\
Sequence & Count 	& Index & Symbol & Sequence & Count & Index & Symbol \\ 
\midrule
1 	& 64 	& 1, 2, 3, ... 32 		&5m-64p & 9 	    & 64 	& 1, 2, 3, ... 32   	&10m-64p\\
2 	& 32 	& 1, 3, 5, ... 31		&5m-32p & 10 	& 32 	& 1, 3, 5, ... 31	    &10m-32p\\
3 	& 16 	& 1, 5, 9, ... 29 		&5m-16p	& 11 	& 16 	& 3, 7, 11, ... 31 	&10m-16p\\
4 	& 8 		& 1, 9, 17, 25 		&5m-8p  & 12 	& 8 	    & 3, 11, 19, 27	    &10m-8p \\
5 	& 8	 	& 2, 10, 18, 26 	&5m-8p  & 13 	& 8 	    & 4, 12, 20, 28	    &10m-8p\\
6 	& 16 	& 2, 6, 10, ... 30 	&5m-16p& 14 	& 16 	& 4, 8, 12, ... 32	&10m-16p\\
7 	& 32		& 2, 4, 6, ... 32 		&5m-32p& 15 	& 32 	& 2, 4, 6, ... 32	    &10m-32p\\
8 	& 64 	& 1, 2, 3, ... 32		&5m-64p& 16 	& 64 	& 1, 2, 3, ... 32   	&10m-64p\\
\bottomrule
\end{tabular}
\end{table}

The experiments were conducted on two circles of radii of 5m and 10 m, respectively. For each circle, experiments with four pedestrian counts (8, 16, 32, 64) were carried out, each of which was repeated four times. With these specific number of participants (e.g., $8 = 64/2^3$), it is convenient to formulate a symmetric experimental situation for all the participants based on the existing 64 surface marks in practice. The detailed personnel schedule arrangement is presented in Table. \ref{tab1}. The schedule arrangement not only avoids an identical initialization condition but also guarantees a time-saving implementation. Note that each participant number in Table. \ref{tab1} respectively corresponds to two pedestrians in A group and B group. All the 16 sequences in Table. \ref{tab1} was performed two times, with one experiment from a semicircle to the opposite semicircle, and the other experiment going back to the original position. Besides, a small number of warm-up experiments were performed for the participants to get familiar with the experiment rules and the related objects, e.g., surface marks.

A high-definition camera operating at 25 frames per second, was placed at a high-rise building beside the public square to record the whole experimental square in an approximate top-down view (see Fig. \ref{figcirclegamec}). In the video process, we applied the color recognition mode of software PeTrack \citep{Boltes2010,Boltes2013} to track the locations of heads (actually the colorful hats), and then calculate their ground positions according to the pedestrian height and the view angle. It is worth noting that the special features in our experiments, e.g., the approximate top-down view and the color requirements of clothes and hats, benefit the precise recognition of the pedestrian trajectories at each frame, even in a crowded situation. In addition, more videos about the experiments can be found on our website, \url{http://pedynamic.com/circle-antipode-experiments/}.

\section{Experiment analysis}\label{section 3}
% 1. 数学定义。
% 2. Individual behavior 

\begin{figure}[!ht]
\centering\includegraphics[width=17cm]{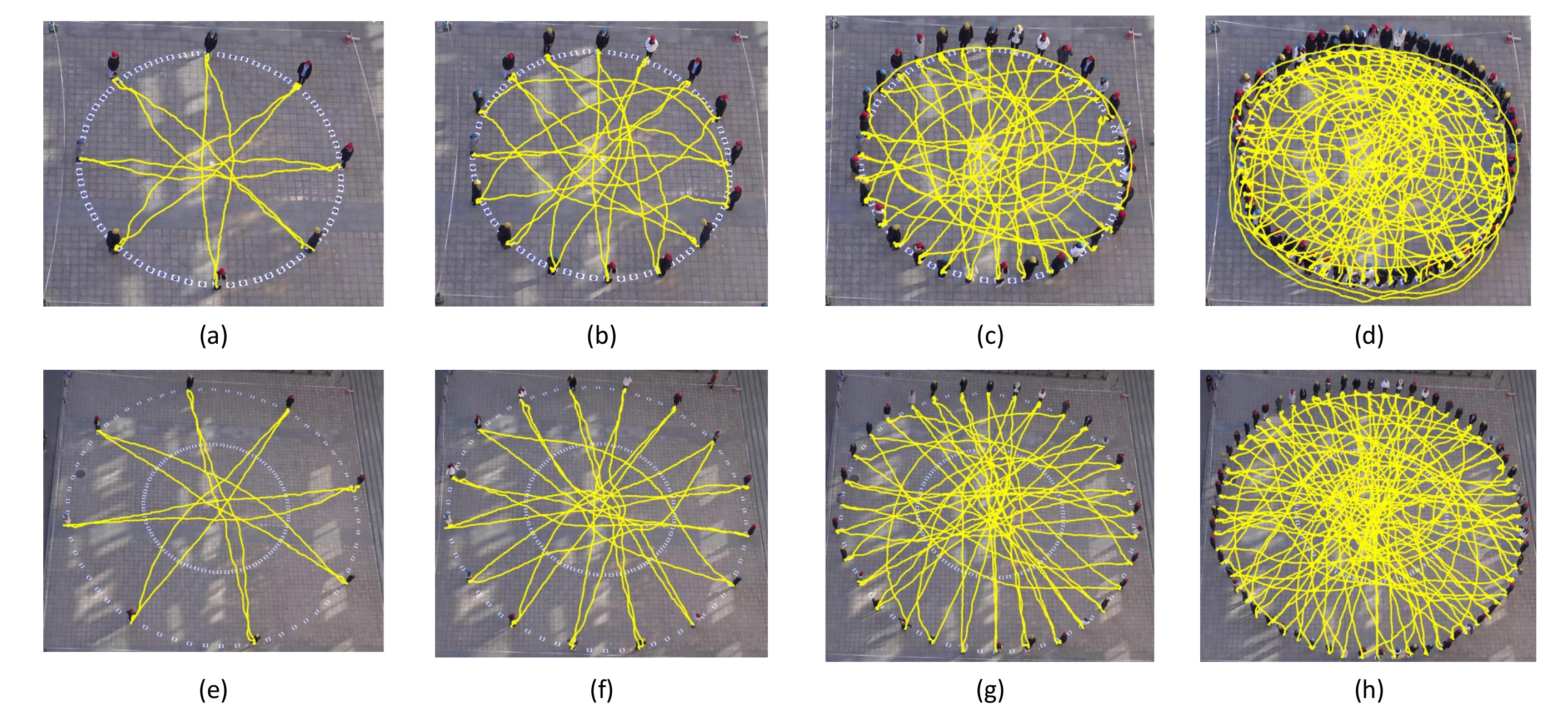}
\caption{Snapshots of the circle antipode experiments and the original trajectories. The yellow lines indicate the corresponding ground trajectories. (a) - (d) contain the 5m experiments with 8, 16, 32, 64 pedestrians, respectively. (e) - (h) contain the 10m experiments with 8, 16, 32, 64 pedestrians, respectively.} \label{figtrajectories}
\end{figure}

In general, the experiments were carried out smoothly, as well as the extraction of pedestrian trajectories. Template original trajectories for the 8 different experiment types are illustrated in Fig. \ref{figtrajectories}. For the sake of convenience, a few abbreviations are introduced to represent the experiments with the same features. The 8 types of experiments are respectively denoted as 5m-8p experiments, 5m-16p experiments, 5m-32p experiments, 5m-64p experiments, 10m-8p experiments, 10m-16p experiments, 10m-32p experiments, and 10m-64p experiments (see Table. \ref{tab1}). Besides, the 5m (10m) experiments include all the experiments in the circle of radius 5(10) meters, and the 8p(16p/32p/64p) experiments include all the experiments with just 8(16/32/64) participants.

\begin{figure}[!ht]
\centering\includegraphics[width=6 cm]{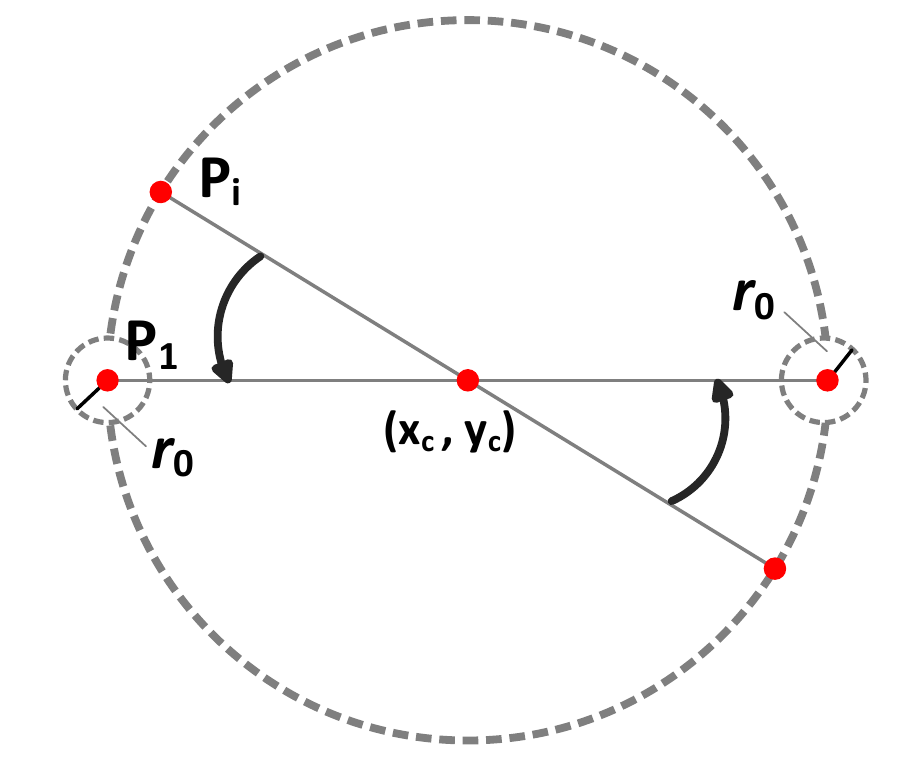}
\caption{Sketch map of the pedestrian trajectory rotation in the circle antipode experiments.} \label{figtrajectories0}
\end{figure}

In the section, the original trajectories extracted from the video constitute the data source for the quantitative analysis. Here, each trajectory is represented with two labels: the pedestrian index and the time step. For the experiment with $N$ pedestrians, the original trajectory is represented as, 
\begin{equation}\label{eq original tra}
\bm{s}_i(t) =\{(x_i(t), \, y_i(t)) \ | \ 1 \leq i \leq N,\ t_i^{start} \leq t \leq t_i^{dest}, \ i,\, t\, \in \mathbb{Z} \}, 
\end{equation}
where $i$ and $t$ are the corresponding pedestrian index and time step, respectively. $x_i(t)$ and $y_i(t)$ represent the coordinates of pedestrian $P_i$ at time step $t$. $t_i^{start}$ and $t_i^{dest}$ denote the departure time from the starting point and the arrival time at the destination point for pedestrian $P_i$, respectively. 
% 可能这边的因素还是得说清楚一些，因为十分重要！！
Considering the swaying phenomenon and other errors \citep{Grieve1966,Kim2004,Hoogendoorn2005,Jelic2012}, two cut-off circles of $r_0$ are respectively provided for both the starting point and the destination point, and $r_0 = 0.5 \rm{m}$ (see Fig. \ref{figtrajectories0}). The departure time $t_i^{start}$ is obtained as the time instant when pedestrian $P_i$ firstly leaves the cut-off circle of the starting point, and the arrival time $t_i^{dest}$ is derived when pedestrian $P_i$ firstly reaches the cut-off circle of the destination point.

\begin{figure}[!ht]
\centering\includegraphics[width=17cm]{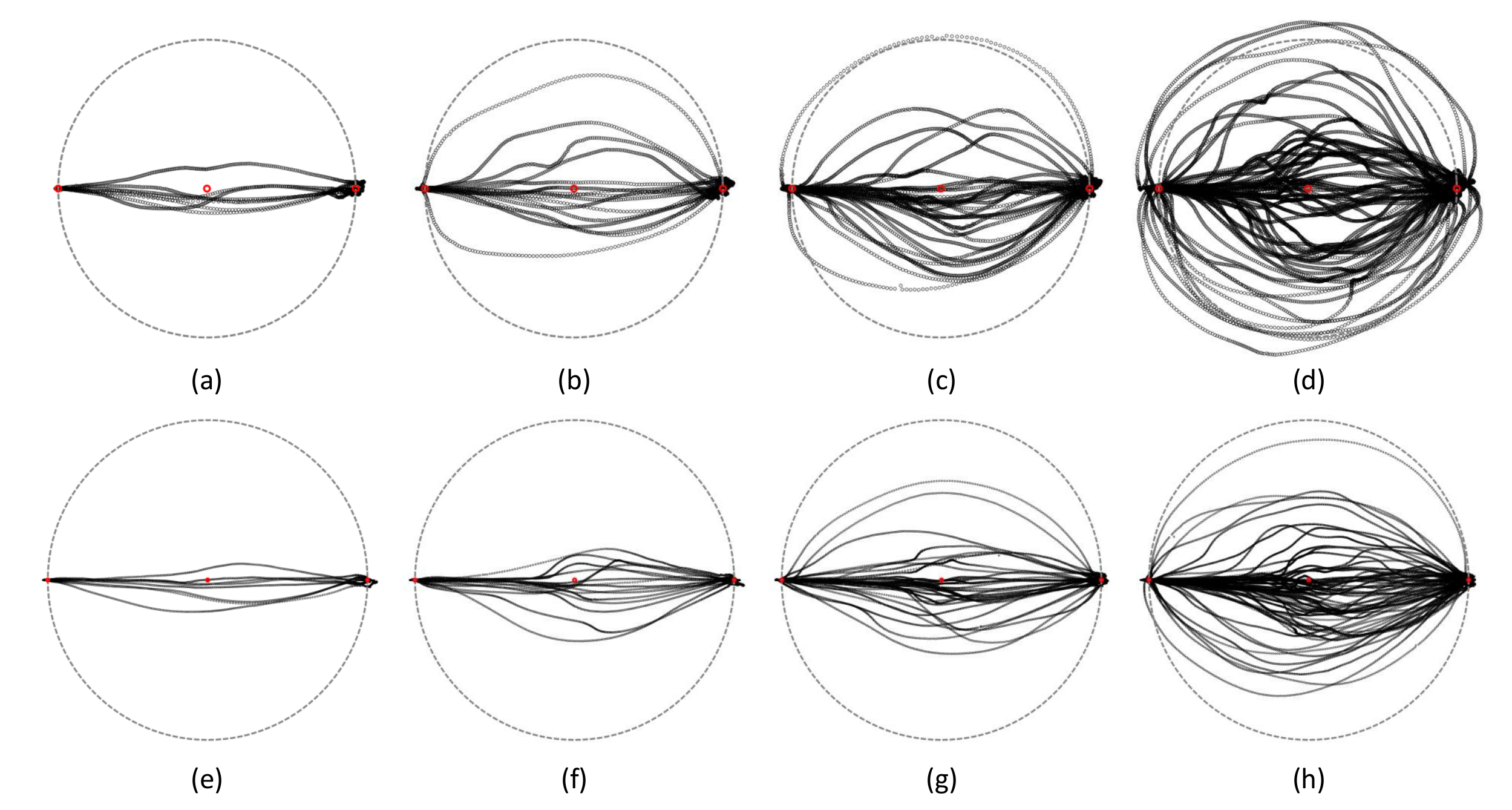}
\caption{Illustrations of the rotated pedestrian trajectories. The black points stand for the rotated trajectories, and the gray circle is an auxiliary line representing the initial circle. The red points from left to right are the starting point, center point and destination point, respectively. (a) - (d) show the 5m experiments with 8, 16, 32, 64 pedestrians, respectively. (e) - (h) show the 10m experiments with 8, 16, 32, 64 pedestrians, respectively.} \label{figtrarotation}
\end{figure}

Moreover, to make the most of the specific symmetric distribution features, the original route trajectories are respectively rotated around the center of the circle until the starting point of the route and the left end of the circle are overlapped (as shown in Fig. \ref{figtrajectories0}). Since the destination of each pedestrian locates at the antipodal position of the starting point, the destination should also be consistent. Suppose the center of the circle is $(x^c, \ y^c)$, the rotation formula of the trajectory $\bm{s}_i(t)$ for pedestrian $P_i$ can be expressed as, 
\begin{equation}	\label{eq rotation}
\left \{
\begin{array}{r}
x_i^R (t) = x^c + (x_i (t)-x^c) \cos ( \frac {2\pi(i-1)} {N}) - (y_i (t) - y^c) \sin( \frac {2\pi(i-1)} {N}) \\
y_i^R (t) = y^c + (x_i (t)-x^c) \sin ( \frac {2\pi(i-1)} {N}) + (y_i (t) - y^c) \cos( \frac {2\pi(i-1)} {N}) \\
\end{array}
\right. .
\end{equation}
Accordingly, the rotated trajectory is represented as, 
\begin{equation}\label{eq original trarot}
\bm{s}_i^R(t)=\{(x_i^R(t), \, y_i^R(t)) \ | \ 1 \leq i \leq N,\ t_i^{start} \leq t \leq t_i^{dest}, \ i,\, t\, \in \mathbb{Z} \}, 
\end{equation}
where $x_i^R(t)$ and $y_i^R(t)$ are the rotated coordinates of pedestrian $P_i$ at time $t$. After the rotation, the starting points and the destination points are consistent according to the practical rotated routes, and the shape of the rotated trajectories is similar to the spindle apparatus in cells from an intuitive view. Consequently, the pedestrian routes can be estimated and investigated through a more comparable perspective (see Fig. \ref{figtrarotation}).

From a qualitative view of the original and rotated trajectories (see Fig. \ref{figtrajectories} and Fig. \ref{figtrarotation}), the individual routes are nearly smooth, and radical changes in the routes are rare. That's to say, even for the circle antipode experiments with serious conflicts between pedestrians, the velocity adjustments are generally gradual in practice. Another significant trend is that the overall detour level just increases with the growing count of pedestrians. It makes sense since a growing number of pedestrians leads to a more crowded environment in the center region, and more pedestrians have to detour around and further increase the overall detour level. 

The following subsections concern the quantitative trajectory discussions, including trajectory spatial distribution, route length, route potential, travel time, speed, and time-series parameters.

\subsection{Trajectory spatial distribution}\label{section spatial distribution}

\begin{figure}[!ht]
\centering 
%\hspace{-0ex} 
%\vspace{-6ex}
\subfloat[The 5m experiments.]{ \label{figdiscentera}\includegraphics[width=0.4\textwidth]{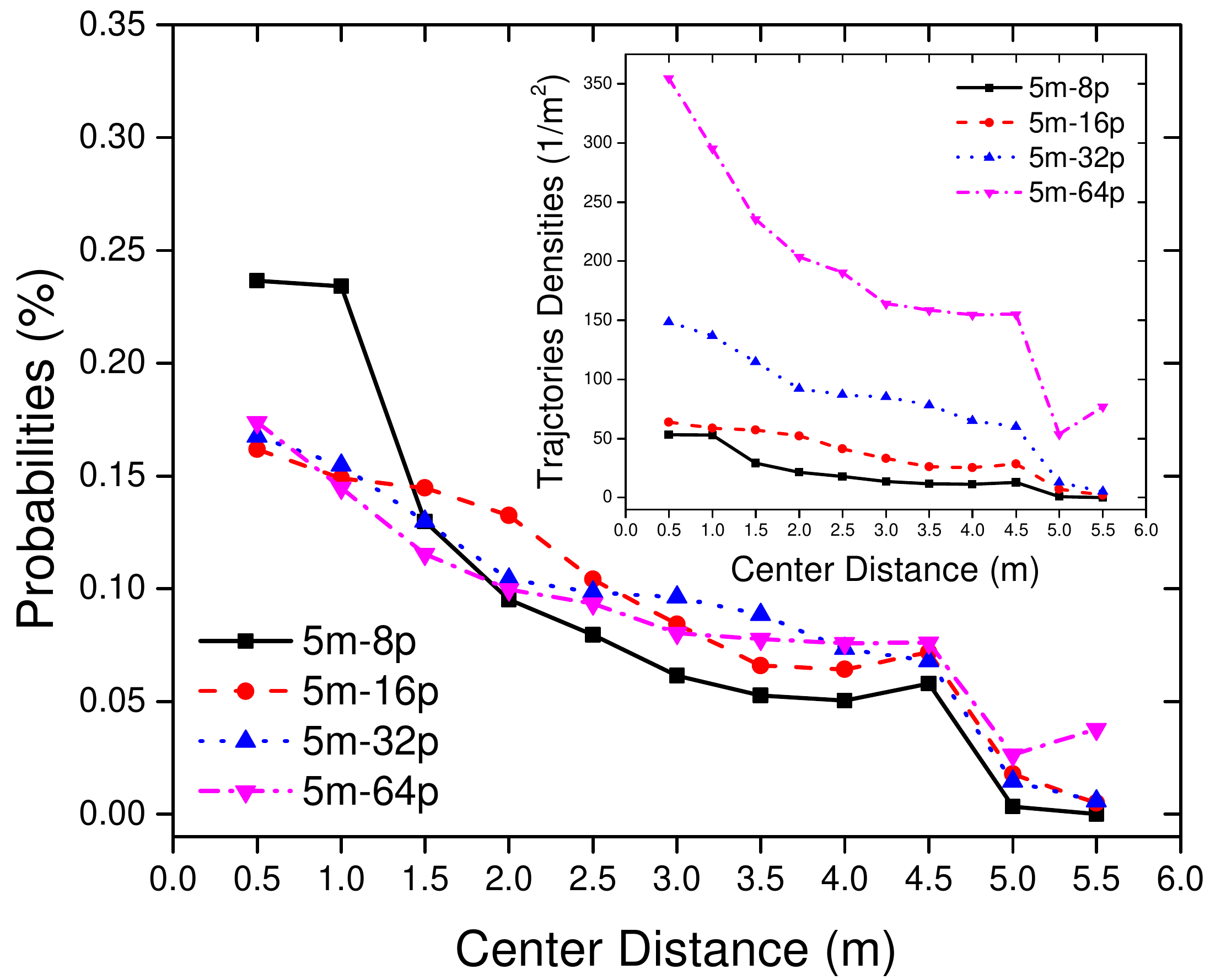}}
%\hspace{-6ex}
\subfloat[The 10m experiments.]{ \label{figdiscenterb}\includegraphics[width=0.4\textwidth]{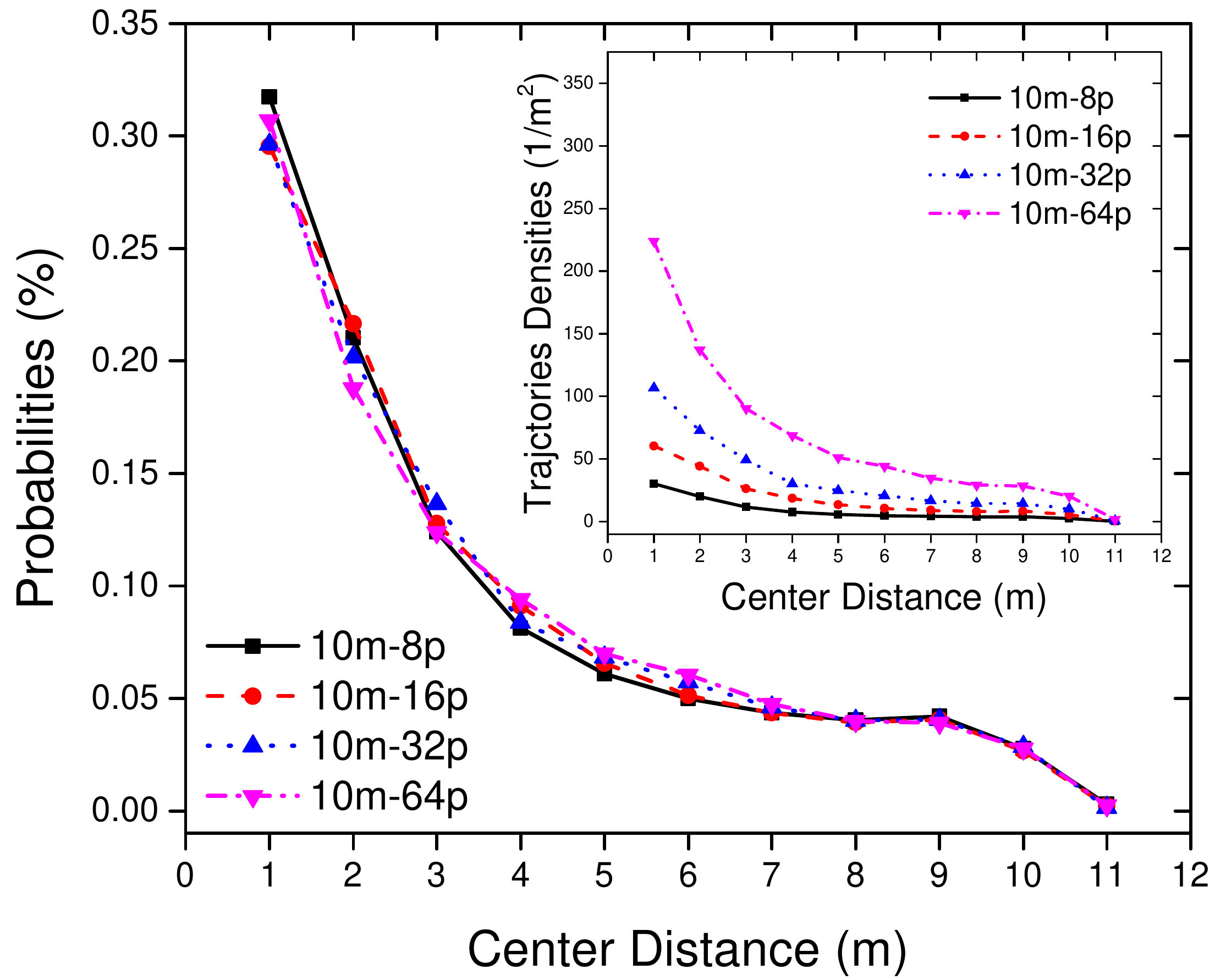}}
\caption{The center distance based spatial distribution of trajectories.} \label{figdiscenter} 
\end{figure}

In the circle antipode experiments, the shortest routes intersect at the center of the circle, and the circle center just locates at the midpoints of the shortest routes. Considering the specific position of the center point $(x^c, y^c)$, the center distance $d_i^c(t)$ is introduced and defined as,
\begin{equation}
d_i^c(t) = \sqrt{ (x_i(t) - x^{\rm{c}})^2 +(y_i(t) - y^{\rm{c}})^2 }.
\end{equation} 
% The trajectories are classified into $N_c +1 $ groups according to the center distance, 
% \begin{equation}\label{eq center zone}
% \bm{s}_i(t) \in 
% \left \{
% \begin{array}{lc}
% \bm{C}_i, & {\rm if} \quad \frac{i-1}{N_c} \cdot r \leq d_i^c(t) < \frac{i}{N_c} \cdot r, \ 1 \leq i \leq N_c, i \in \mathbb{Z}\\
% \bm{C}_{N_c+1}, & \rm{Otherwise}. 
% \end{array}
% \right. ,
% \end{equation}
Besides, the pedestrian trajectories are discrete in our base data, and the densities of trajectories are calculated as follows:
\begin{equation}
\rho(d) =\frac {C(d)} {\pi d^2 - \pi(d-\Delta d)^2},
\end{equation}
where $C(d)$ is the count of trajectories within the center distance range $(d-\Delta d, d)$, and $\Delta d$ equals to 0.5 meters and 1 meter in the 5m and 10m experiments, respectively. Fig. \ref{figdiscenter} illustrates the center distance based trajectory densities and probabilities in the experiments. In general, the trajectory density goes down along with the growing center distance. The trend can be explained according to the special experiment setting that the shortest routes of pedestrians just intersect at the center of the circle. Hence, along with the approaching of the circle center, the area is going to be more competitive which leads to the negative correlation between the center distance and the trajectories density. Also, the trajectories density increases with the growing of pedestrian counts, and it can be account for the growing congestion caused by more pedestrians. The figure also shows that the trajectories in the 10m experiments are more concentrated in the center region than that in the 5m experiments. In our view, the rise of the distance to the center region in the 10m experiments provides more room for pedestrians to adjust the conflict avoidance strategies crossing the center region, whereas more pedestrians in the 5m experiments have to detour. In the two types of circles, the changing trends of probabilities are approximately the same, that is to say, the experiments share similar motion patterns. Still, the probabilities of trajectories in the range $d_i^c(t) > 5\rm{m}$ are greater in the 5m-64p experiments than those in the other 5m experiments. Note that the trajectories ($d_i^c(t) > 5\rm{m}$) located outside the 5 m circle are supposed to be detour behaviors, and they are believed to be caused by the overcrowded situation in the 5m-64p experiments. 

% 上升，下降 
%总的来看，行人的轨迹密度是随着center distance 的增加不断下降的。这主要是因为在该实验中，行人的最短路径穿越圆心到达圆的对角位置。而圆心也是行人最短路径的交点，因此随着距离中心位置的靠近，行人的轨迹会更加密集。 同时我们发现，随着实验中行人人数的增加，中心区域的轨迹密度也是不断变大。这一点是很好理解的，主要是因为人数增加后，行人的拥挤度提高了。除此之外，通过观察center distance 的 概率密度可以发现，10m 实验中行人轨迹的相对分布在中心区域更加集中。这一点是因为5m实验中，行人的缓冲距离更小，也因此在中心区域更为拥堵，行人会更加倾向于选择绕行，从而导致了5m实验中行人的轨迹分布更加不集中。两类圆中中不同实验的概率密度变化趋势基本一致，但是在 5m-8p的实验中，行人的轨迹密度在离中心点较近的距离处比例及较大，这可能是因为该实验条件下，基本没有出现拥堵，所以行人基本都是从中心点穿越的。另外，5m-64p 实验中，行人的轨迹密度比例在 dis > 5m时 明显较大。值得注意的是，dis>5m时，行人的轨迹实际上已经处于圈外了。这主要是因为在该实验中，行人过度拥挤，所以有部分行人会采取圈外绕行的行为所导致的。

\begin{figure}[!ht]
\centering 
%\hspace{-0ex} 
%\vspace{-6ex}
\subfloat[The 5m experiments.]{ \label{figquadranta}\includegraphics[width=0.4\textwidth]{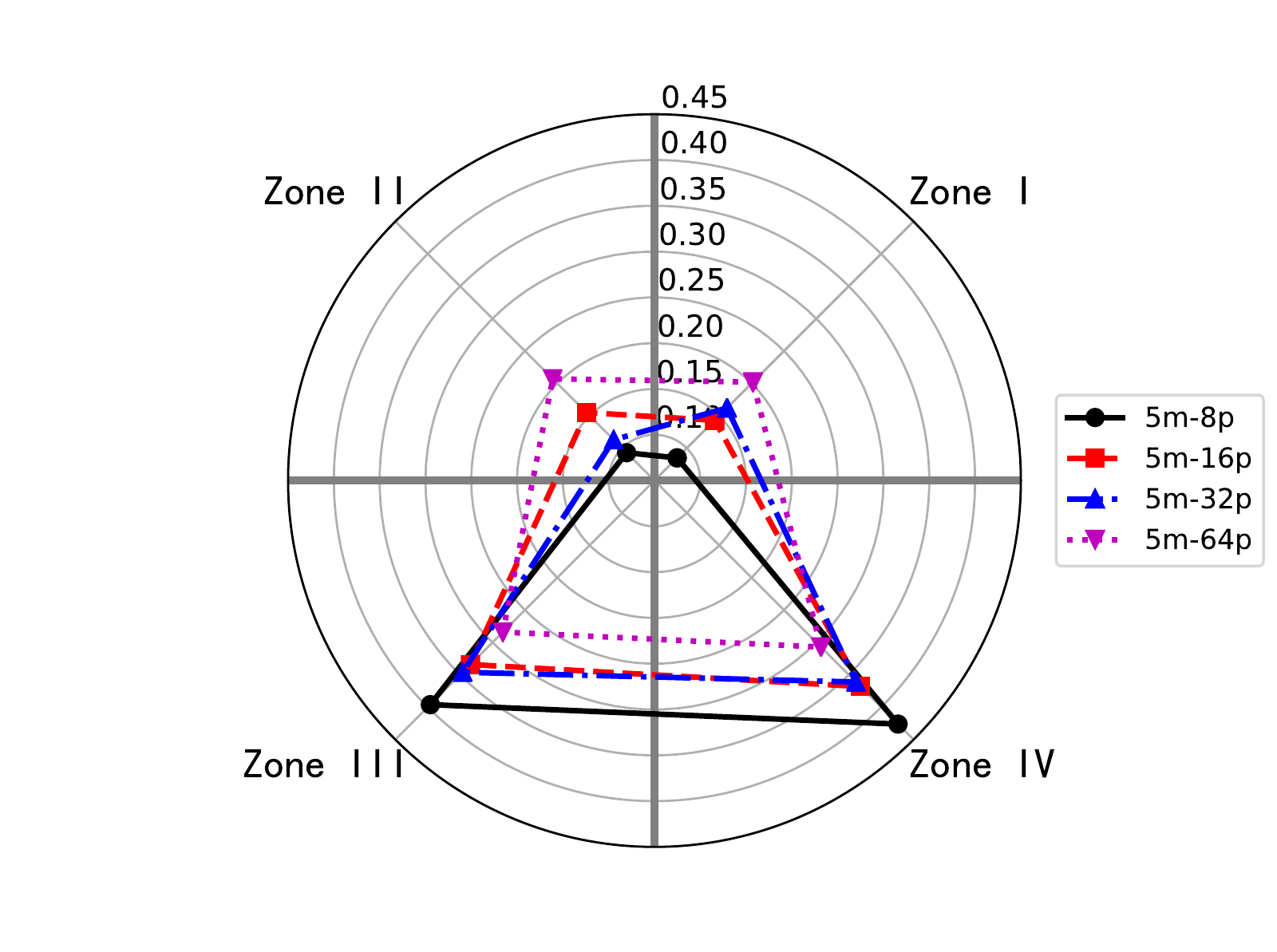}}
%\hspace{-6ex}
\subfloat[The 10m experiments.]{ \label{figquadrantb}\includegraphics[width=0.4\textwidth]{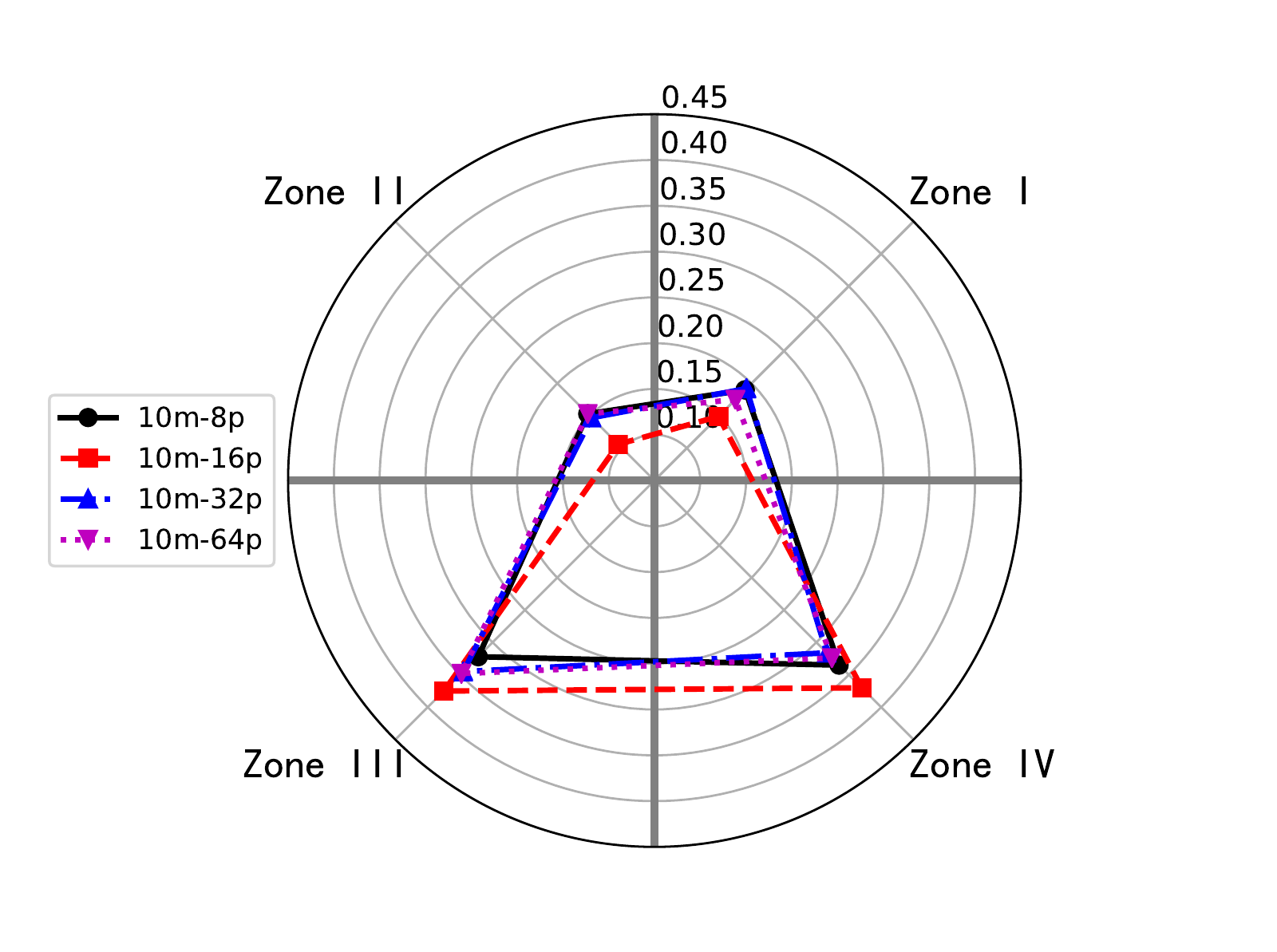}}
\caption{The quadrant circle based spatial distribution of rotated trajectories.} \label{figquadrant} 
\end{figure}

The rotation of pedestrian routes (Fig. \ref{figtrarotation}) provides a more apparent view of the behavior characteristics. Generally, the detour behaviors become more frequent as the pedestrian counts increase, and the phenomenon performs more obvious in the 5m experiments. The pedestrians in the 10m experiments have more space and buffer time for the adjustments on velocity including both direction and speed. Consequently, the congestion in the center region is less likely to be formulated, and the detour level is lower in the 10m experiments.

For a further analysis of spatial distribution of the rotated trajectories, the circle is divided into four zones as shown in Fig. \ref{figtrarotation}. The top-right quadrant, top-left quadrant, bottom-left quadrant, and bottom-right quadrant are defined as Zone I, Zone II, Zone III, and Zone IV, respectively. In the circle, the coordinate of the center point is denoted as $(x^c, y^c)$, and the trajectories are classified into different zones by, 
\begin{equation}\label{eq zone}
\bm{s}_i^R(t) \in 
\left \{
\begin{array}{lr}
Z_1, & {\rm if} \ x_i^R(t) \geq x^c, y_i^R(t) \geq y^c \\
Z_2, & {\rm if} \ x_i^R(t) < x^c, y_i^R(t) \geq y^c \\
Z_3, & {\rm if} \ x_i^R(t) < x^c, y_i^R(t) < y^c \\
Z_4, & {\rm if} \ x_i^R(t) \geq x^c, y_i^R(t) < y^c
\end{array}
\right. ,
\end{equation}
where $Z_1$, $Z_2$, $Z_3$, and $Z_4$ denote the trajectories set of Zone I, Zone II, Zone III, and Zone IV, respectively. For each experiment set, the average ratio of trajectories in each zone is presented in Fig. \ref{figquadrant}. Note that the trajectories within a $r_0 = 0.5{\rm m}$ radius of the original point and the destination point are removed from the dataset to ensure the accuracy of the analysis. It is found that the probabilities of trajectories in Zone III and Zone IV are greater than those in Zone I and Zone II. That is to say, pedestrians prefer to detour from the right side in our experiments. Past studies \citep{Helbing2005,Moussaid2009} suggested that the side preference could be interpreted as a cultural bias, and the right side preference in China agrees with our experimental results. Besides, the probabilities of trajectories in Zone I and Zone IV are approximately equal to those in Zone II and Zone III. Namely, approximate time is spent in the first and the last semicircle.

\subsection{Route length}\label{section length}

The route length of pedestrian $P_i$ is calculated based on the individual trajectories, i.e., 
\begin{equation}\label{eq routelength}
L_i = \sum_{t=t_i^{\rm{start}}}^{t_i^{\rm{dest}}-1} \| s_i(t+1)-s_i(t) \| + 2r_0.
\end{equation} 
Note that $2r_0$ is introduced to compensate for the impact of the cut-off circles, and the analyzed video in the experiments is 25 frames per second. Moreover, the additional length $L^a$ is introduced to compare different experiments. For each circle, the shortest route lengths $L^0$ of the pedestrians are the same and equal to $2r$, pointing from the starting point to the destination point. Based on the feature, the additional length of pedestrian $P_i$ is defined as, 
\begin{equation} \label{eq additional length}
L_i^a = L_i - L_i^0.
\end{equation}

\begin{figure}[!ht]
\centering 
\includegraphics[width=0.4\textwidth]{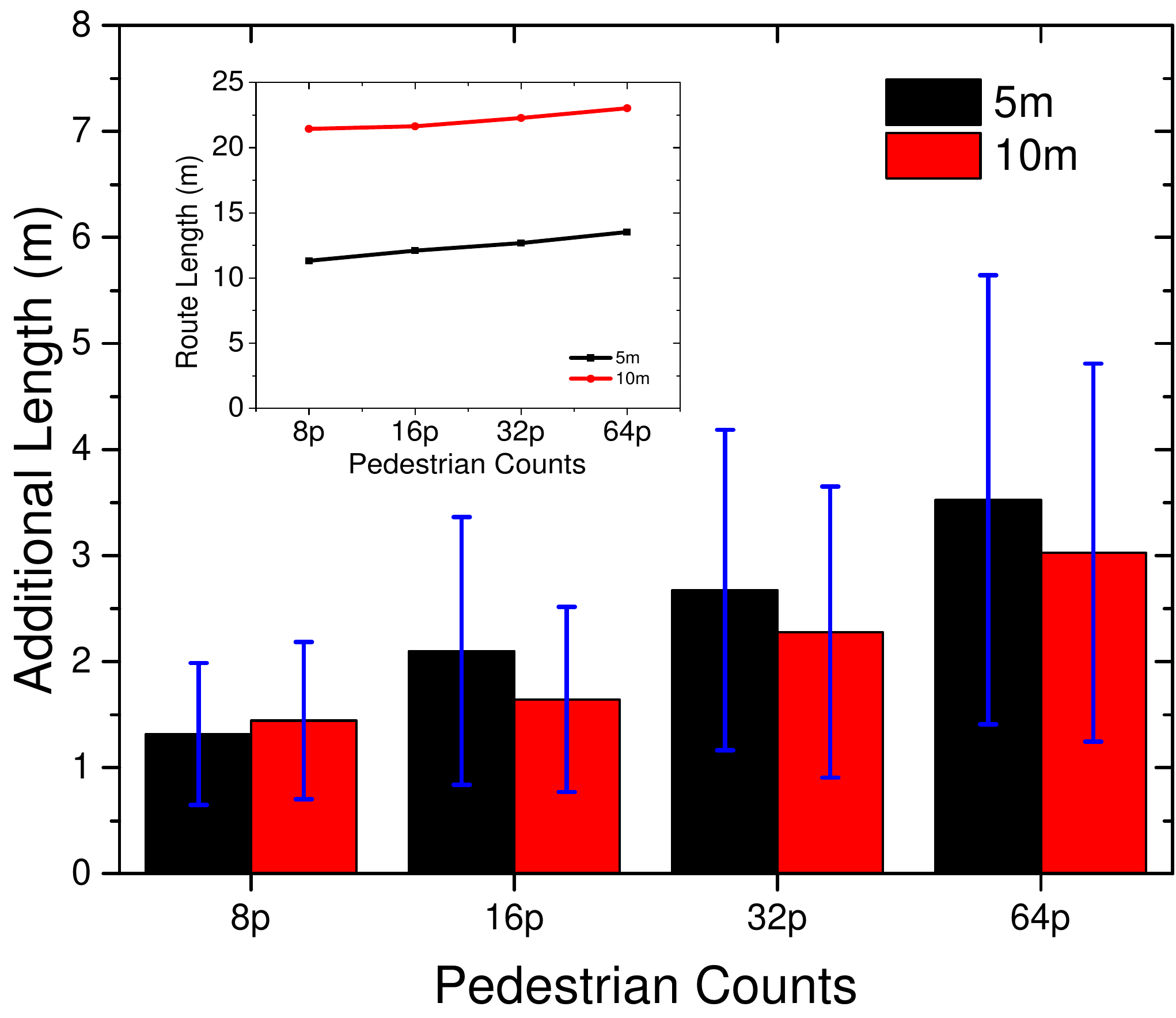}
\caption{The mean values and standard deviations of additional length in the experiments.} \label{figlength} 
\end{figure}

% Since 2000, however, the cost of risk capital has gone up in spite of dramatically falling interest rates.
% As the banks' earnings rise, the proportion that has to be paid out in dividends for Huijin to finance repayments drops.
% With the household income of middle-class Americans falling because of declining wages or growing medical costs, parents are having trouble paying for their kids and taking care of their own parents
Fig. \ref{figlength} summaries the mean values and the standard deviations for the additional length in the experiments. In the 5m and the 10m experiments, the additional length increases with the growing of pedestrian counts. It makes sense that a more congested center region is likely to formulate with the increased pedestrian counts, and more pedestrians choose to take longer routes. The comparisons between the 5m experiments and the 10m experiments show that, except for the 8p experiments, the additional lengths in the 5m experiments are greater than those in the 10m experiments. To understand the differences, we have to figure out the two components of the additional length. The first component of the additional length comes from the actual detour distance determined by the congestion level, and a pedestrian is more likely to take a longer detour route under a more congested situation. The second component is caused by the swaying phenomenon and other errors \citep{Grieve1966,Kim2004,Hoogendoorn2005,Jelic2012}. For instance, even in a completely undisturbed scenario, the practical route lengths are larger than the shortest ones. This component is continuously accumulated along with the movement distance. Accordingly, in the 8p experiments, the conflicts among pedestrians are not enough to cause a jam in the center region, and the pedestrians are likely to choose the approximate shortest routes. Hence, the first component of additional length is similar between the 5m-8p experiment and the 10m-8p experiment. In the case, the second additional length component plays a major role in formulating the differences between the two types of experiments, and the mean additional length in the 10m-8p experiments is greater. In the 16p, 32p, and 64p experiments, the congestion regions in the center region are more serious, and pedestrians need to walk a longer distance to reach their destinations, so the first component of additional length component outperforms. In this case, a greater congestion area is generally formulated in the 5m experiment and the pedestrians have to detour more. Therefore, the additional length in the 5m experiments is greater than those in the 16p, 32p, and 64p experiments. Moreover, the standard deviation shows the similar trend as average route length, and it can be explained in the same way.

\begin{figure}[!ht]
\centering 
\subfloat[The 5m experiments.] { \label{figdislengtha} \includegraphics[width=0.4\textwidth]{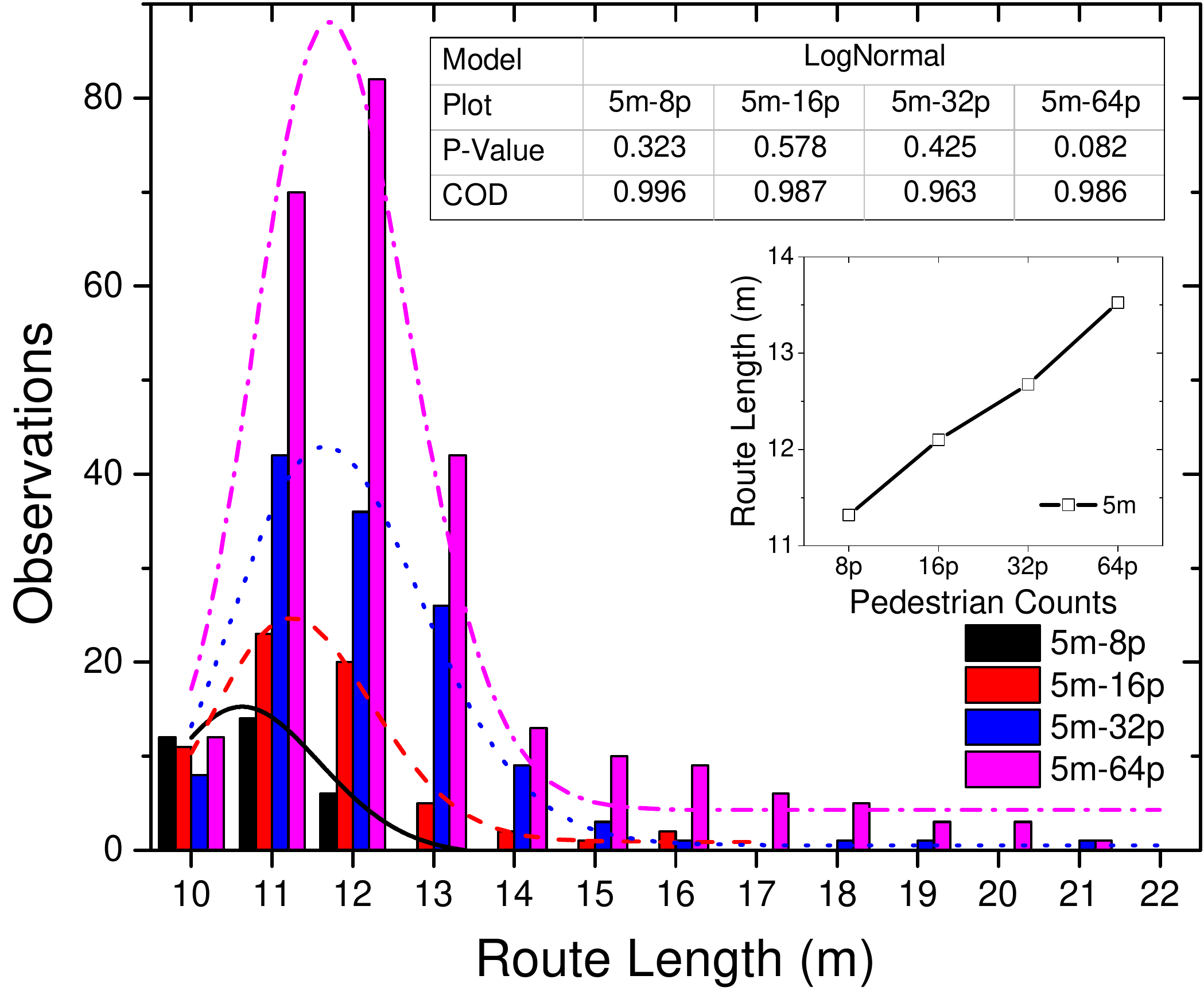}}
\subfloat[The 10m experiments.]{ \label{figdislengthb} \includegraphics[width=0.4\textwidth]{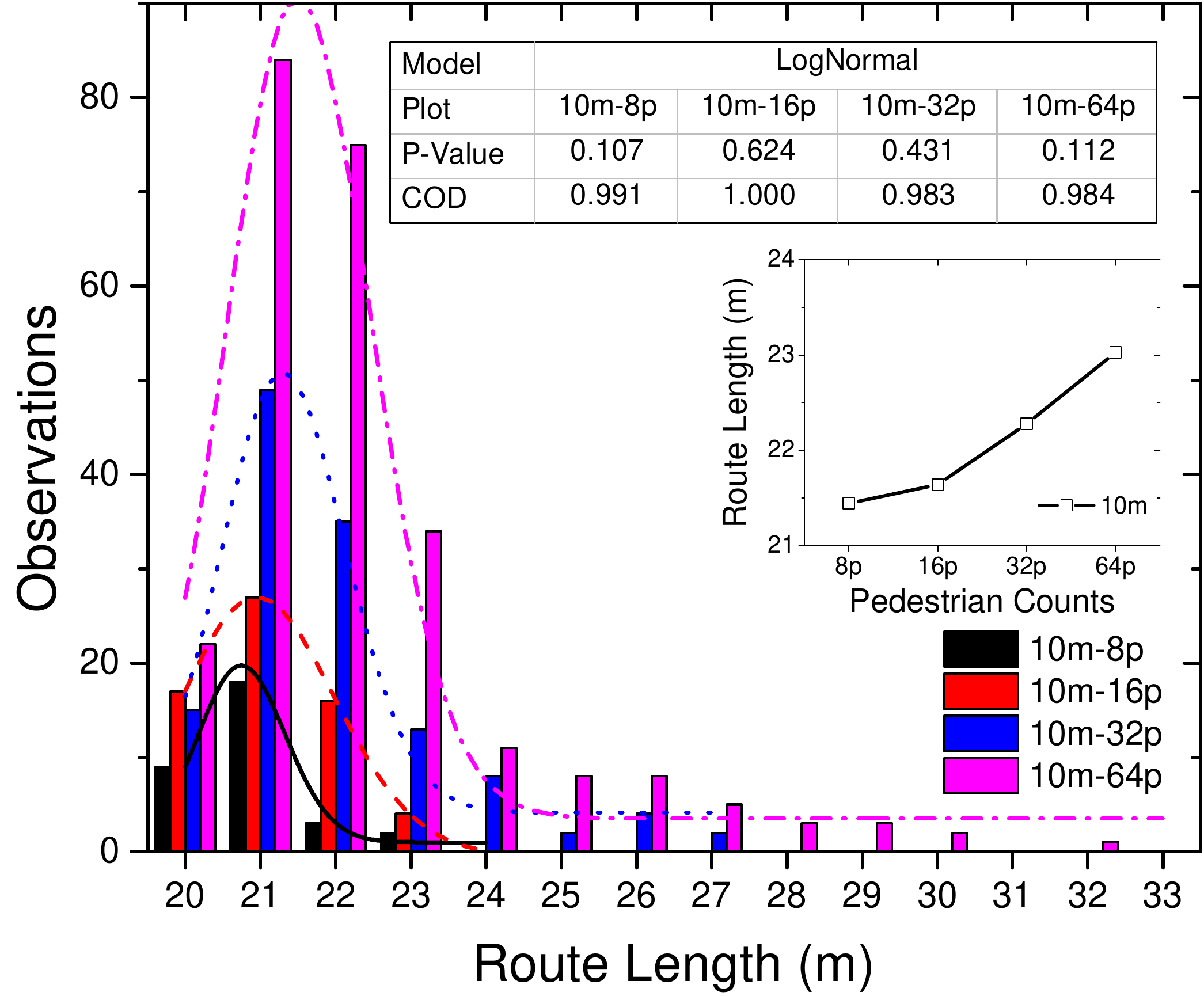}} 
\caption{Distributions of the route lengths.} \label{figdislength} 
\end{figure}

The mean values and the distributions of the route length are illustrated in Fig. \ref{figdislength}. The route lengths cover from 10 meters to 21 meters in the 5m experiments, while from 20 meters to 33 meters in the 10m experiments. On the whole, there is a peak value in each distribution, and the peak value generally exists at the left side of the data range and a long tail exists at the right side. Considering the special shape of the route length distribution, the log-normal distribution is introduced to fit it. In the hypothesis test, the null hypothesis is given as that the route length distribution follows a log-normal distribution, and the significant level is set as 5\%. The results of the non-parametric test show that the p-values in all experiments are greater than 0.05 (see Fig. \ref{figdislength}), so the null hypothesis can not be rejected and the route length obeys a log-normal distribution. Besides, the coefficients of determination (COD) further prove the goodness of fitting.

\subsection{Route Potential} \label{section potential}
\begin{figure}[!ht]
\centering\includegraphics[width=0.4\textwidth]{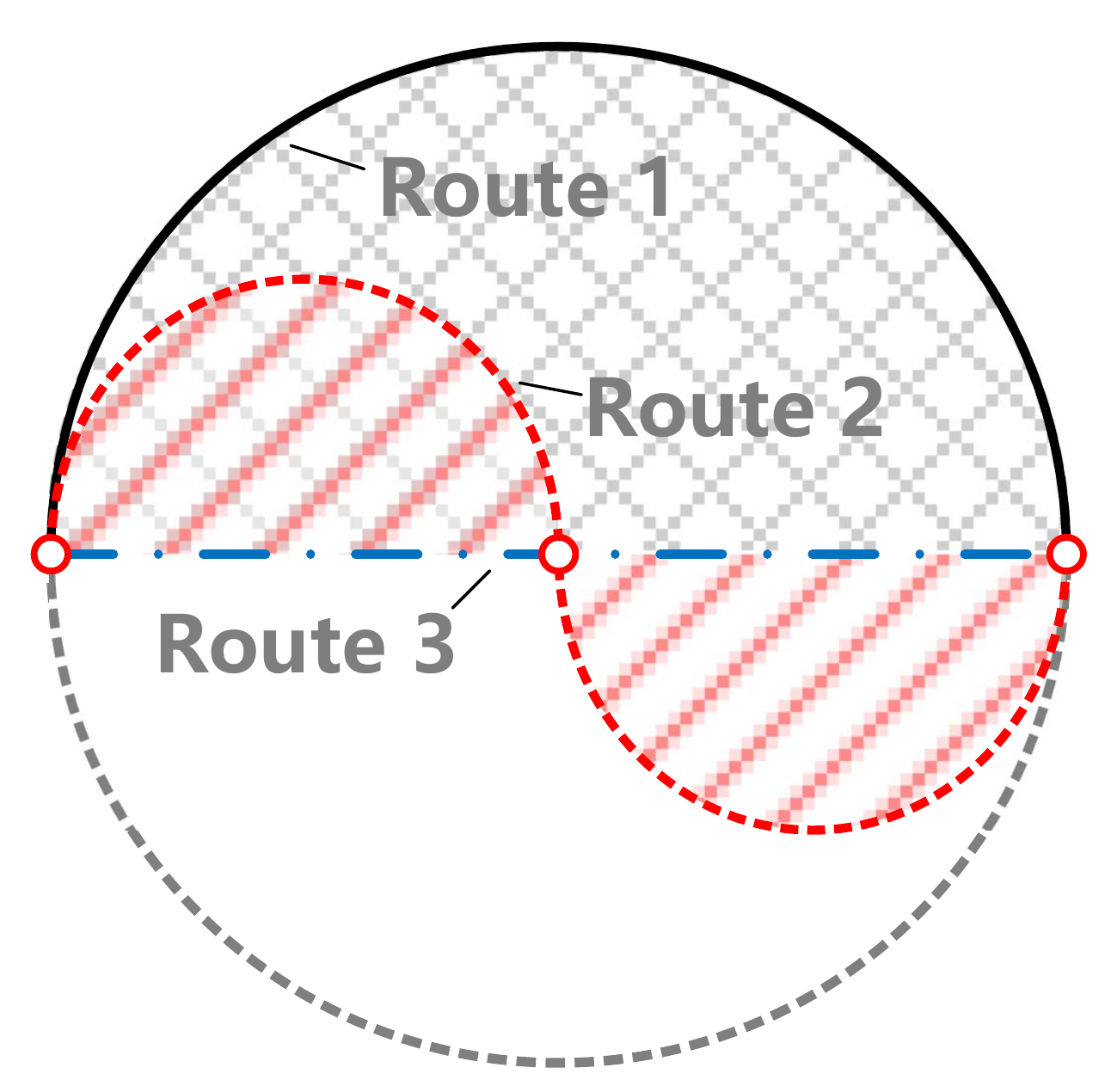}
\caption{Illustration of route length and route potential.} \label{figpotential}
\end{figure}

The route potential is defined as the area surrounded by the practical route and the shortest route in Fig. \ref{figpotential}. Based on the discrete characteristics of the rotated trajectories, the route potential of pedestrian $P_i$ is calculated by,
\begin{equation} \label{eq potential}
M_i = \left|\sum_{t=t_i^{\rm{start}}}^{t_i^{\rm{dest}}-1} \left(\frac {y_i^{\rm{R}}(t+1)+y_i^{\rm{R}}(t)} {2} \cdot (x_i^{\rm{R}}(t+1)-x_i^{\rm{R}}(t)) \right)\right|,
\end{equation}
where $M_i$ is non-negative but $y_i^{\rm{R}}(t)$ has a positive or negative sign. In theory, the route potential measures the deviation level of the practical route.

The introduction of the route potential index provides a novel perspective to understand the spatial feature of pedestrian routes. In Fig. \ref{figpotential}, there are three different template routes. Specifically, Route 1 fully detours around the center region based on the trajectory shape of a semicircle of radius $r$. Route 2 moves along two semicircles of radius $0.5r$. Route 3 represents the shortest route from the starting point to the destination point. In general, the three routes respectively respond to three kinds of pedestrian route choices. Route 1 and Route 2 own the same route lengths but show different route potential. Route 2 and Route 3 own the same route potential, but their route lengths are different. In sum, the combination of the route length and the route potential enhances the recognition of the features of a pedestrian route in the space dimension.

\begin{figure}[!ht]
\centering 
\subfloat[The 5m experiments.] { \label{figdispotentiala} \includegraphics[width=0.4\textwidth]{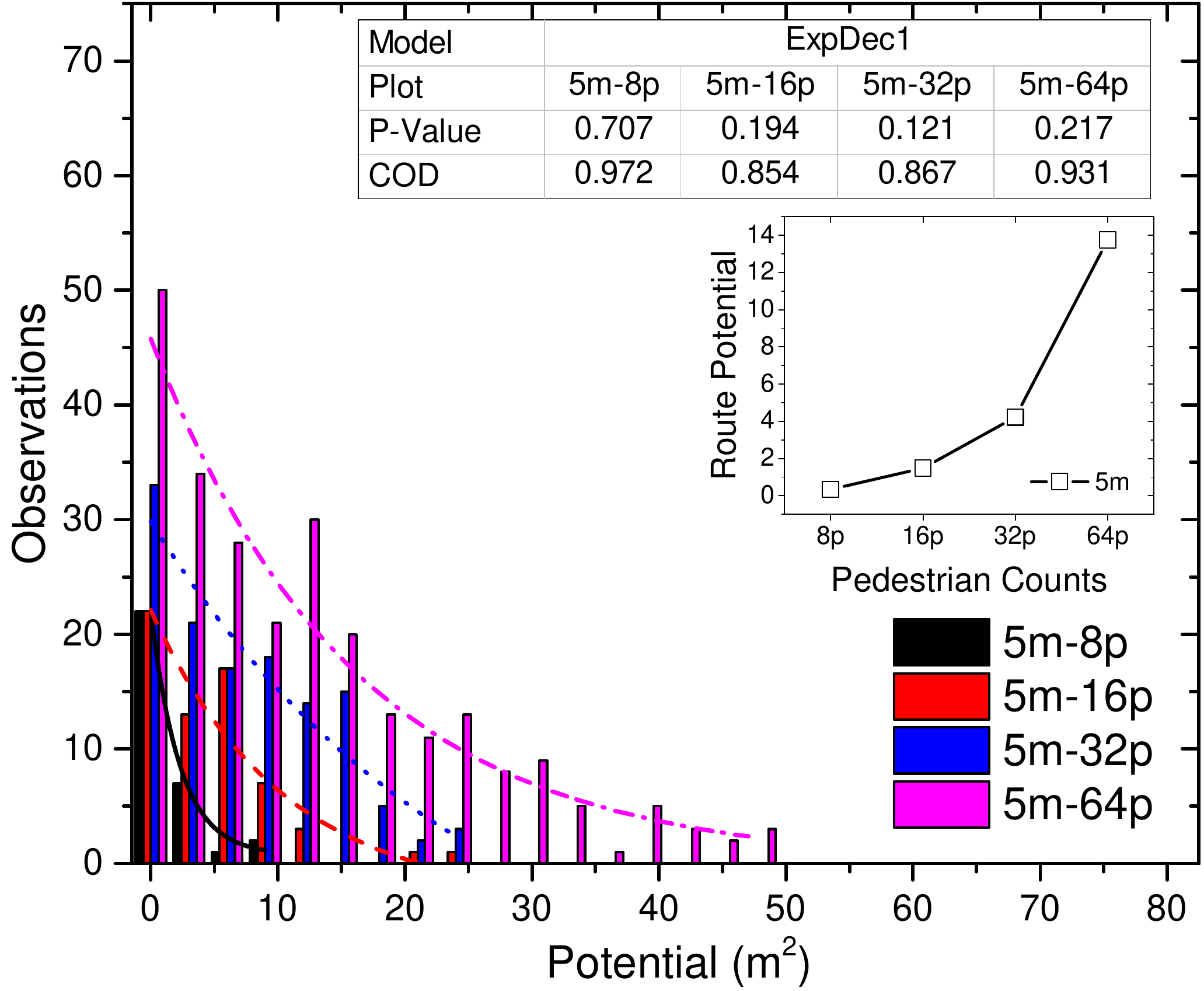}}
\subfloat[The 10m experiments.]{ \label{figdispotentialb} \includegraphics[width=0.4\textwidth]{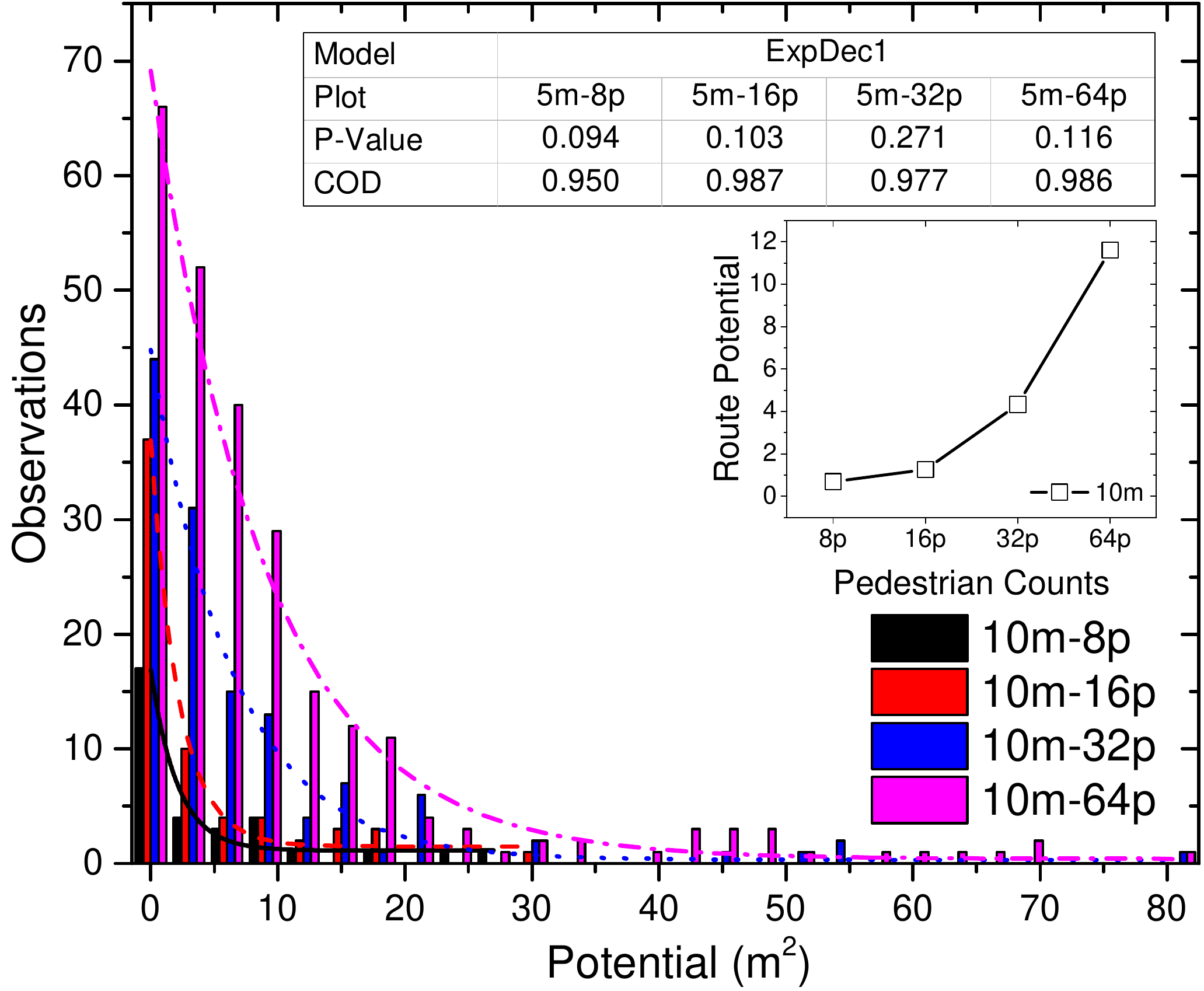}} 
\caption{Distributions of the route potential.} \label{figdispotential} 
\end{figure}

The distributions of the route potential are shown in Fig. \ref{figdispotential}. The mean values of the potential basically grow with the rise of pedestrian counts since more pedestrians have to detour in a more crowded situation. The probability decreases with the growing of route potential, and the exponential distribution is introduced for fitting. The results of the non-parametric test (Fig. \ref{figdispotential}) show that the distributions of the potential agree well with the exponential distributions. In addition, the results of COD still prove the goodness of exponential distribution fitting.

\subsection{Travel time} \label{section time}
% distribution test的说明还是不是很清楚，还是要再思考一下
In the experiments, the travel time of pedestrian $P_i$ is calculated according to the departure time $t_i^{\rm{start}}$ of the cut-off circle from the starting point and the arrival time $t_i^{\rm{dest}}$ of the cut-off circle at the destination point, i.e., 
\begin{equation} \label{eq travel time}
T_i = t_i^{\rm{dest}} - t_i^{\rm{start}}.
\end{equation}

\begin{figure}[!ht]
\centering 
\subfloat[The 5m experiments.] { \label{figdistimea} \includegraphics[width=0.4\textwidth]{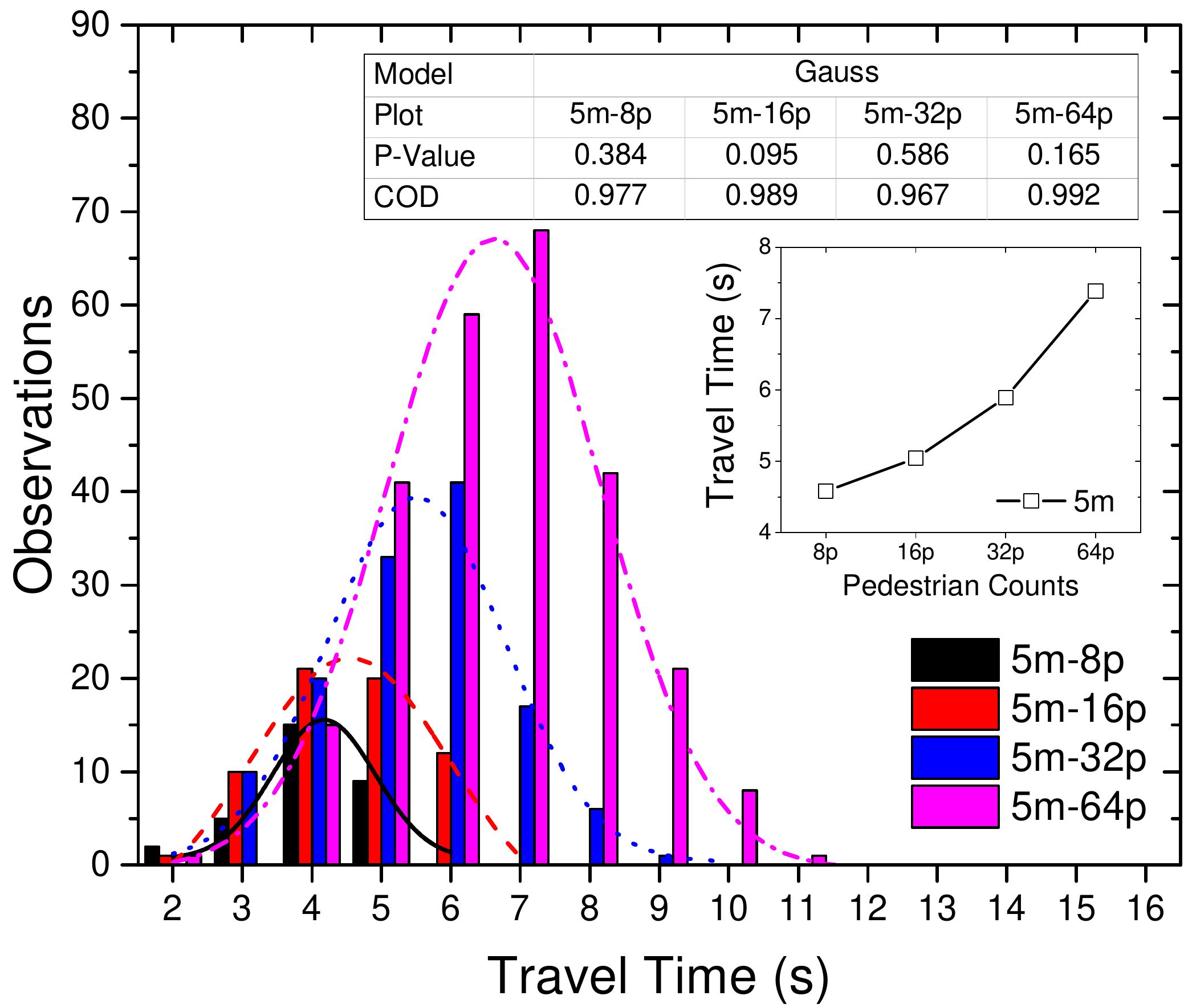}}
\subfloat[The 10m experiments.]{ \label{figdistimeb} \includegraphics[width=0.4\textwidth]{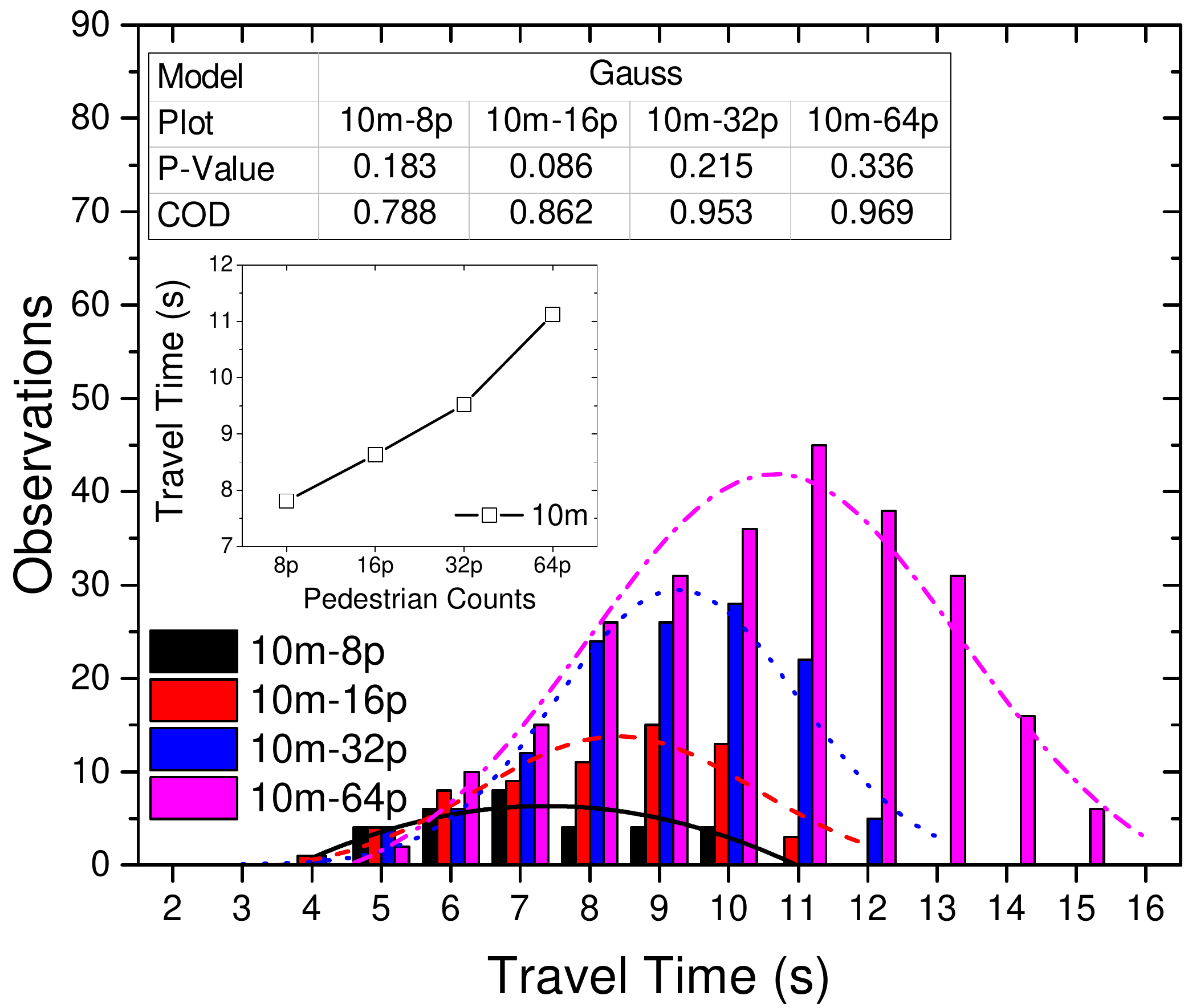}}
\caption{Distributions of the travel time.} \label{figdistime} 
\end{figure}

The mean values and distributions of travel time are illustrated in Fig. \ref{figdistime}. In the experiments, the mean travel time rises with the growing of pedestrian counts. It figures out since the growing pedestrian counts make the circle more crowded and more pedestrians have to detour or slow down, hence the travel time increases. The travel times cover from 2 seconds to 11 seconds in the 5m experiments, while from 3 seconds to 16 seconds in the 10m experiments. Each distribution owns a peak value which increases with the growing of the pedestrian counts. The distributions of time are mountain-liking curves without deflecting to left or right. Hence, the normal(Gauss) distribution is introduced to fit it. The normality tests (Fig. \ref{figdistime}) and the COD prove that the travel time distribution follows a normal distribution.

\subsection{Length-time correlation} \label{section lengthtime}

\begin{figure}[!ht]
\centering \includegraphics[width=0.8\textwidth]{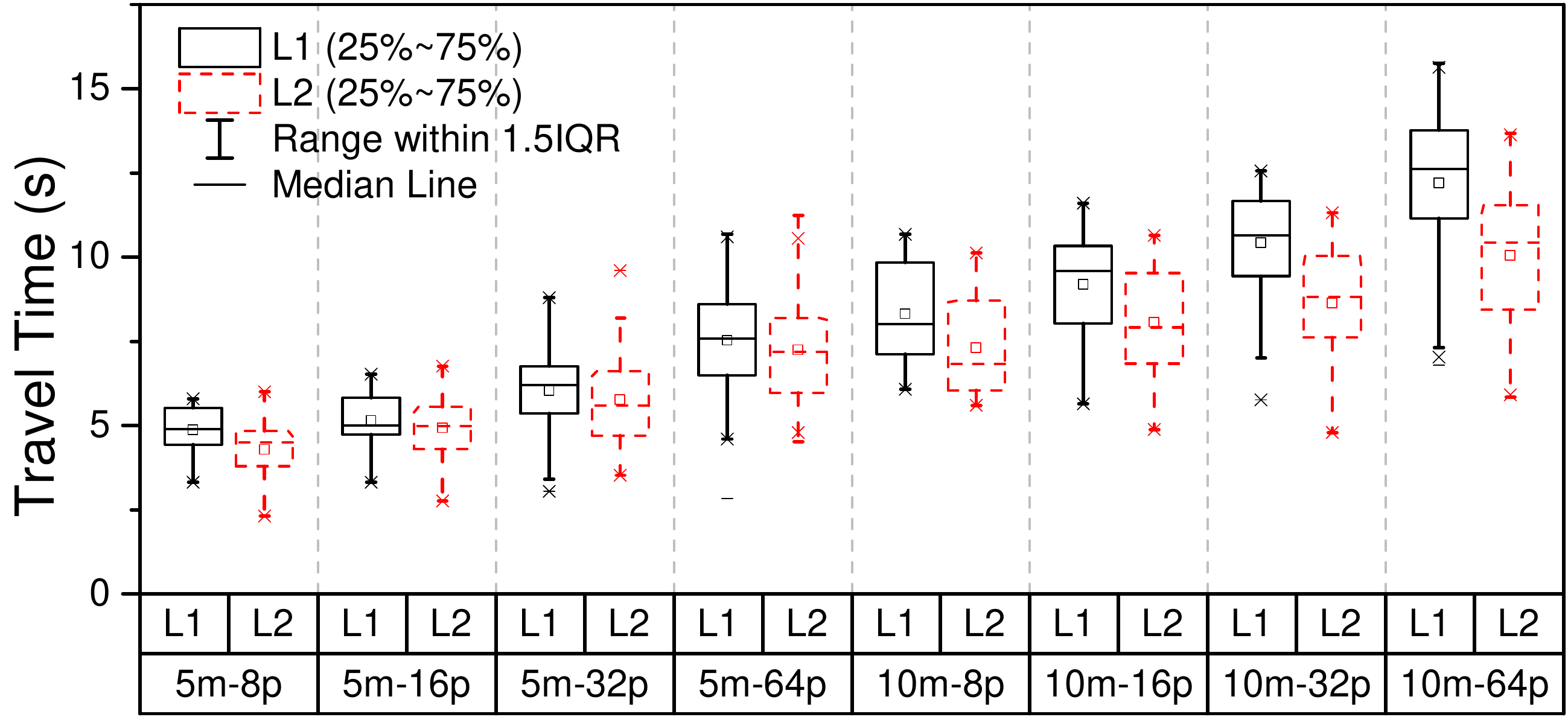}
\caption{Travel time comparisons between the different choices of route length.} \label{figbox} 
\end{figure}

In the experiments, the practical pedestrian routes can be basically divided into two groups. The first type of pedestrians approximately insist on the shortest route and move through the central region, while the second type of pedestrians choose to detour around the central region. Generally, the route length is likely to be the most fundamental indicator to distinguish between the two types of routes. Therefore, the pedestrians are divided into two groups based on the route length to investigate the influence of different route choices. For each experiment, group $L1$ contains the first half of pedestrians with shorter route lengths, while group $L2$ includes the second half with longer route lengths. The travel time of the two groups of pedestrians are represented with the box plots, and the results of the 8 sets of experiments are shown in Fig. \ref{figbox}.

It is found that group $L1$ experiences a greater mean travel time, whereas group $L2$ corresponds to a smaller one. Moreover, the gap between mean travel times exists in all the 8 experiment sets, despite a relatively small one. In summary, the route length and the travel time are not positively associated. This can be demonstrated that most of the pedestrians are attracted to the shortest routes across the center region instead of detouring, which results in crowdedness and further delays pedestrians’ movement. On the contrary, the minority of participants, who avoid the congested area, experience long but time-saving travels. Consequently, the mean travel time of group $L1$ is even greater in the experiments.

The route choice in the circle antipodes experiments is similar to the traffic distribution problem despite the practical pedestrian routes are not fixed. Logically speaking, the user equilibrium (UE) states shall be expected in the experiments, but the practical results show definite differences between the travel time of the two types of route choices. The heterogeneity of pedestrians and the nondeterminacy of behaviors should not be the critical reason since the results are stable in all the 8 sets of experiments. In our view, there are two kinds of causes for the results. First, the participants may not recognize the characteristics of the circle antipode experiments and the status of other participants. As a result, most pedestrians insist on the default shortest route, and more pedestrians choose the shorter routes than those in the UE state. Second, the practical cost function of route choice contains more complicated factors other than travel time. That's to say, reaching the destination as quickly as possible maybe not the only goal for pedestrians, and some other factors such as the energy might be a critical factor as well. In the case, the shorter route is likely to be a more energy-saving choice, and more pedestrians are attracted than those in the UE state.

\subsection {Speed} \label{section speed}
%留下这一部分，但是简要处理。只去分析行人速度大小。

In the section, the speed of pedestrian is calculated according to the original trajectories and denoted by, 
\begin{equation} \label{eq speed}
v_i(t) = \frac { \| \bm{s}_i(t+1) - \bm{s}_i(t) \| } {\Delta{t}},
\end{equation}
where $\Delta{t}$ is the time interval of pedestrian trajectories and it equals to 0.04 s in the experiments.

\begin{figure}[!ht]
\centering 
\subfloat[The 5m experiments.] { \label{figdisvelocitysizea} \includegraphics[width=0.4\textwidth]{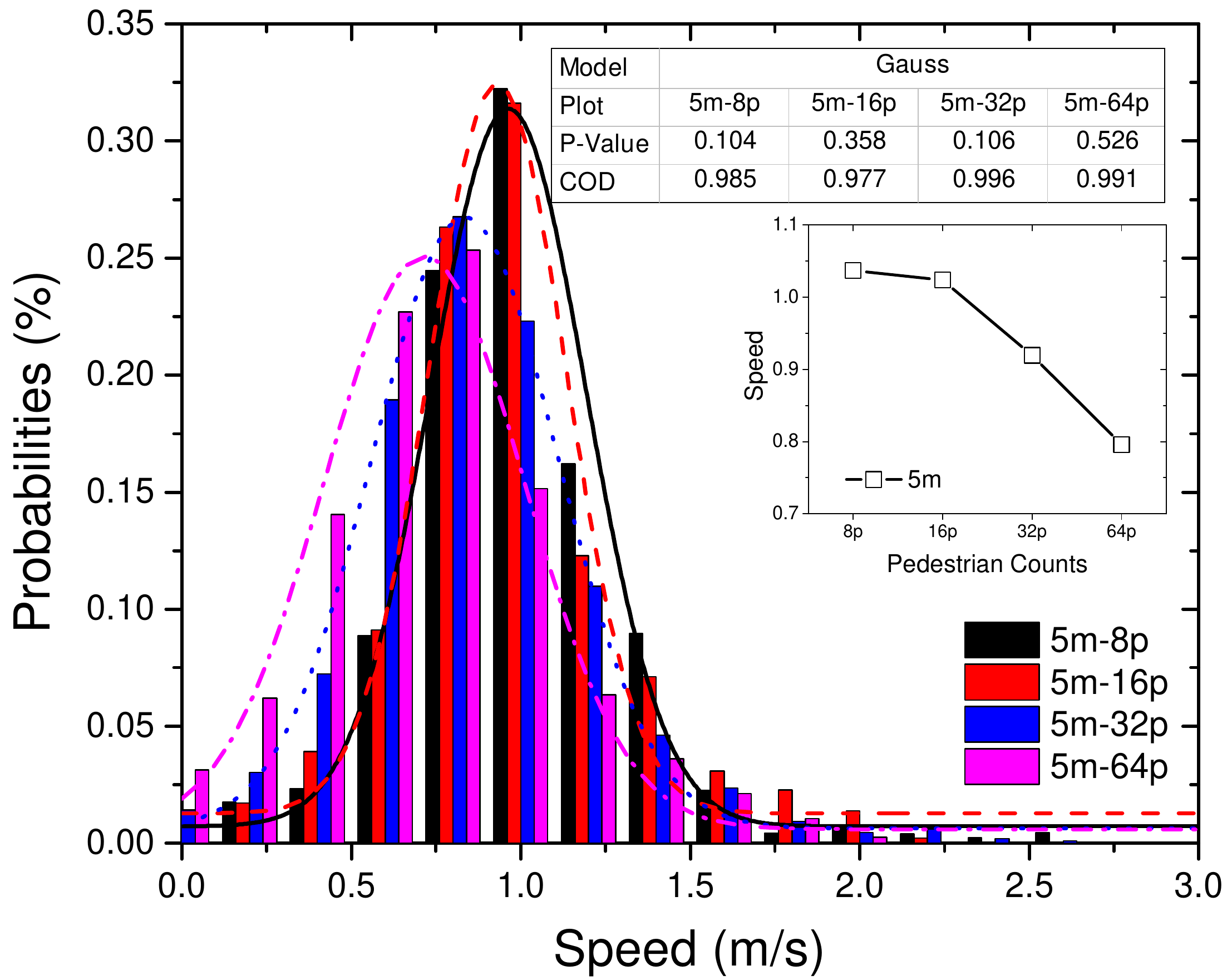}}
\subfloat[The 10m experiments.]{ \label{figdisvelocitysizeb} \includegraphics[width=0.4\textwidth]{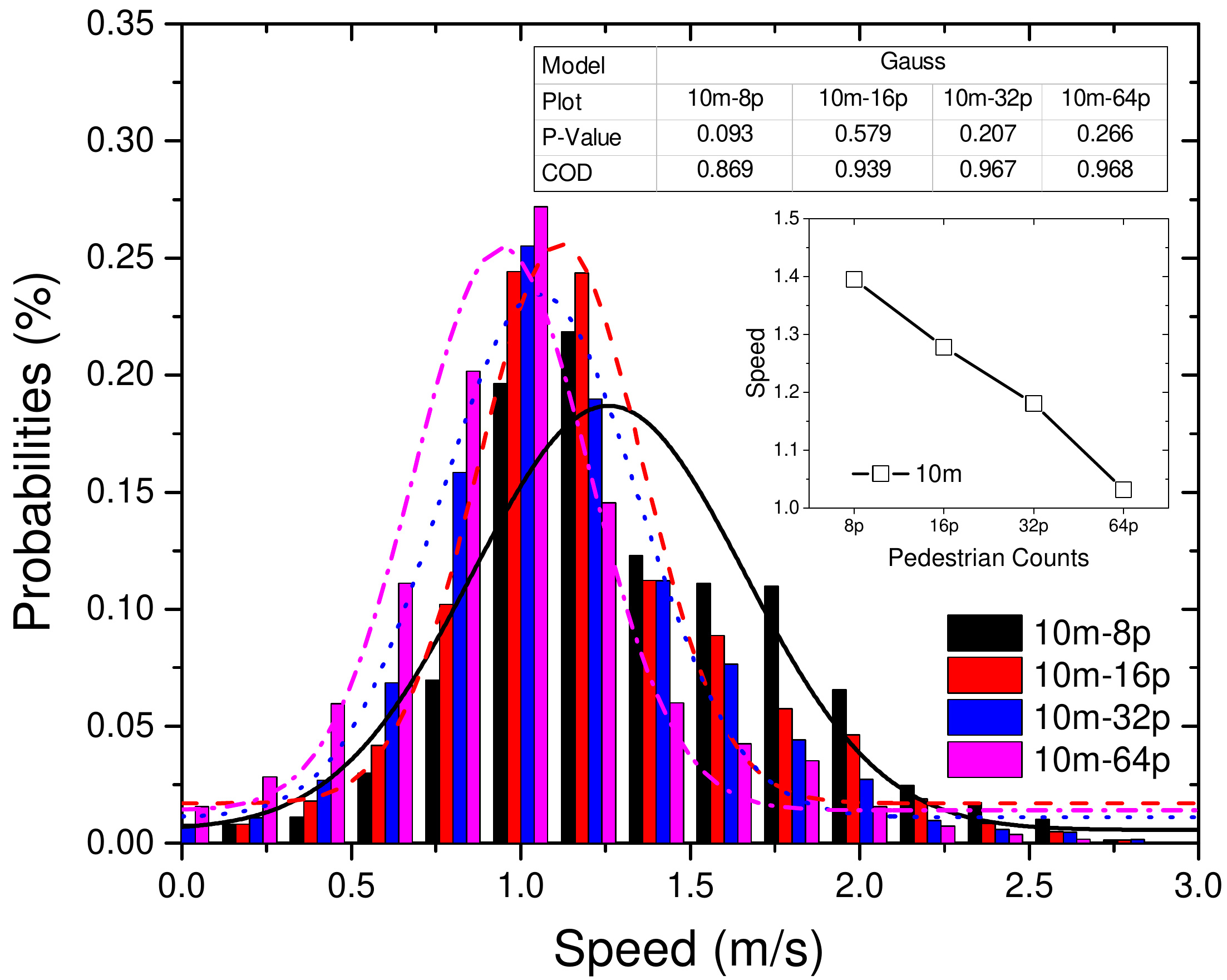}} 
\caption{Distributions of the speed.} \label{figdisvelocitysize} 
\end{figure}

Fig.\ref{figdisvelocitysize} shows the distribution of speed $v_i(t)$ in the series of experiments. The mean velocity drops with the growing of pedestrian counts in the experiments. It makes sense since more pedestrians can result in a more crowded environment and bring down the speed. The peak values of speed occur from 0.75 to 1.25 m/s, which decreases with the growing of pedestrian counts. The speed approximately ranges from 0 to 2 m/s in the 5m experiments, while from 0 to 2.75 m/s in the 10m experiments. The wider motion space in the 10m experiment offers the pedestrians more possibilities to adjust the motion strategies and accelerate to higher speeds. The normality tests and COD in Fig. \ref{figdisvelocitysize} prove that the speed distribution subjecting to a normal distribution cannot be rejected.

\subsection{Time series} \label{section time series}
% 时间序列分析。。。 分析和视频对应。 
% 为什么要做时间序列分析。。。。 因为
The time series indexes can contribute to investigating the status changes of pedestrian crowd during the experiments. For this purpose, the average center distance is introduced as, 
\begin{equation} \label{eq average distance}
\bar{d}^{\rm{c}}(t)= \frac { \sum_{i=1}^{N} \sqrt{ (x_i(t) - x^{\rm{c}})^2 +(y_i(t) - y^{\rm{c}})^2 }} {N}.
\end{equation}
In the experiments, the circle center point is the midpoint and the intersection point of the shortest routes for pedestrians. Considering the specificity of the circle antipode experiment, the average center distance can be used to imply the route choice behaviors of individuals and the aggregation status of crowds. On the individual level, the average center distance measures the distance to the center point, and its changing trend reflects the individual route choice and detour behavior. On the crowd level, the average distance implies the degree of crowdedness. Beyond the average distance, the average velocity is also introduced as, 
\begin{equation} \label{eq average velocity}
\bar{v}(t) = \frac { \sum_{i=1}^{N} v_i(t) } {N}. 
\end{equation}
The average velocity is the other significant index to measure the status of pedestrian crowds. 

\begin{figure}[!ht]
\centering 
\subfloat[The 5m experiments.] { \label{figtimevelocitya} \includegraphics[width=0.4\textwidth]{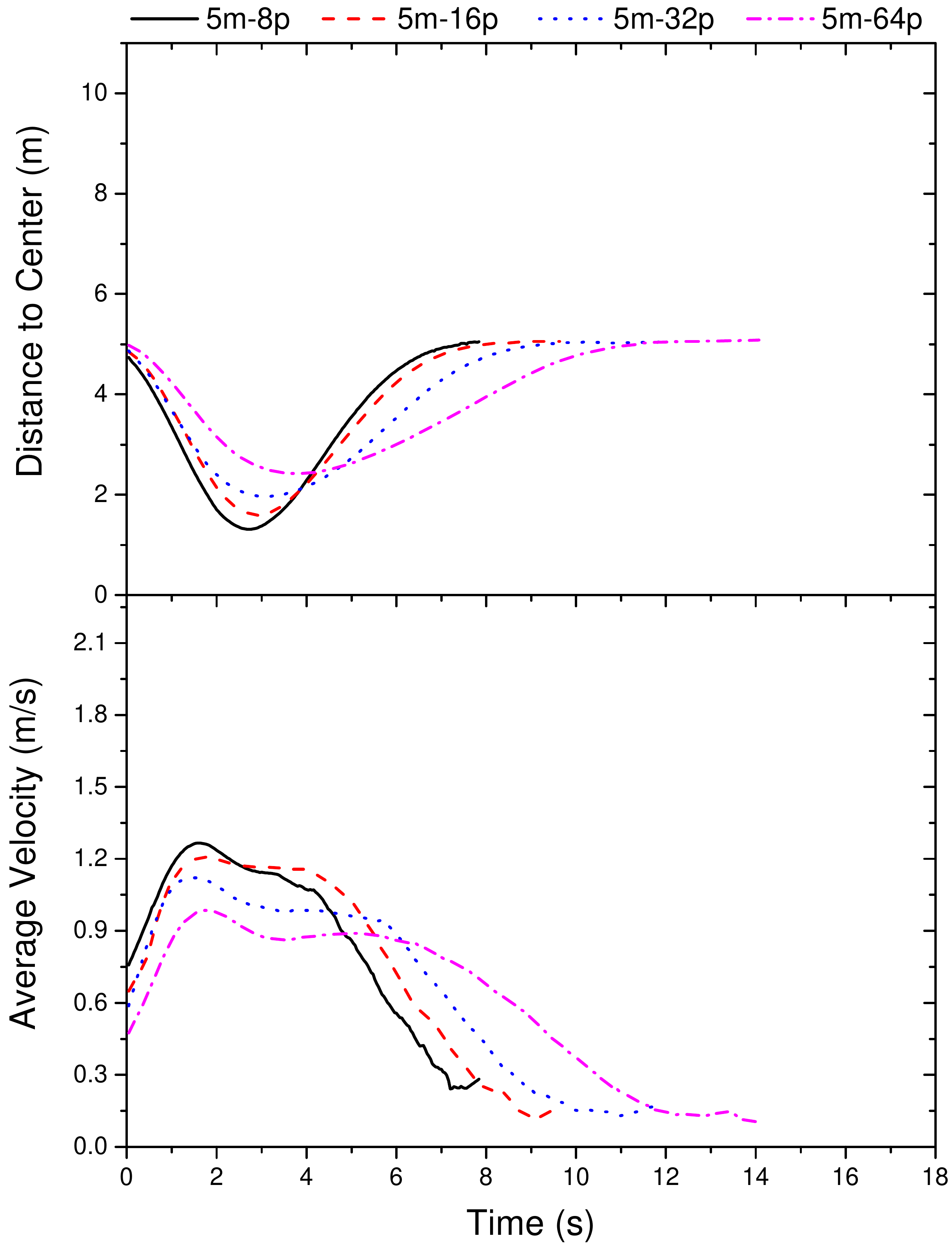}}
\subfloat[The 10m experiments.]{ \label{figtimevelocityb} \includegraphics[width=0.4\textwidth]{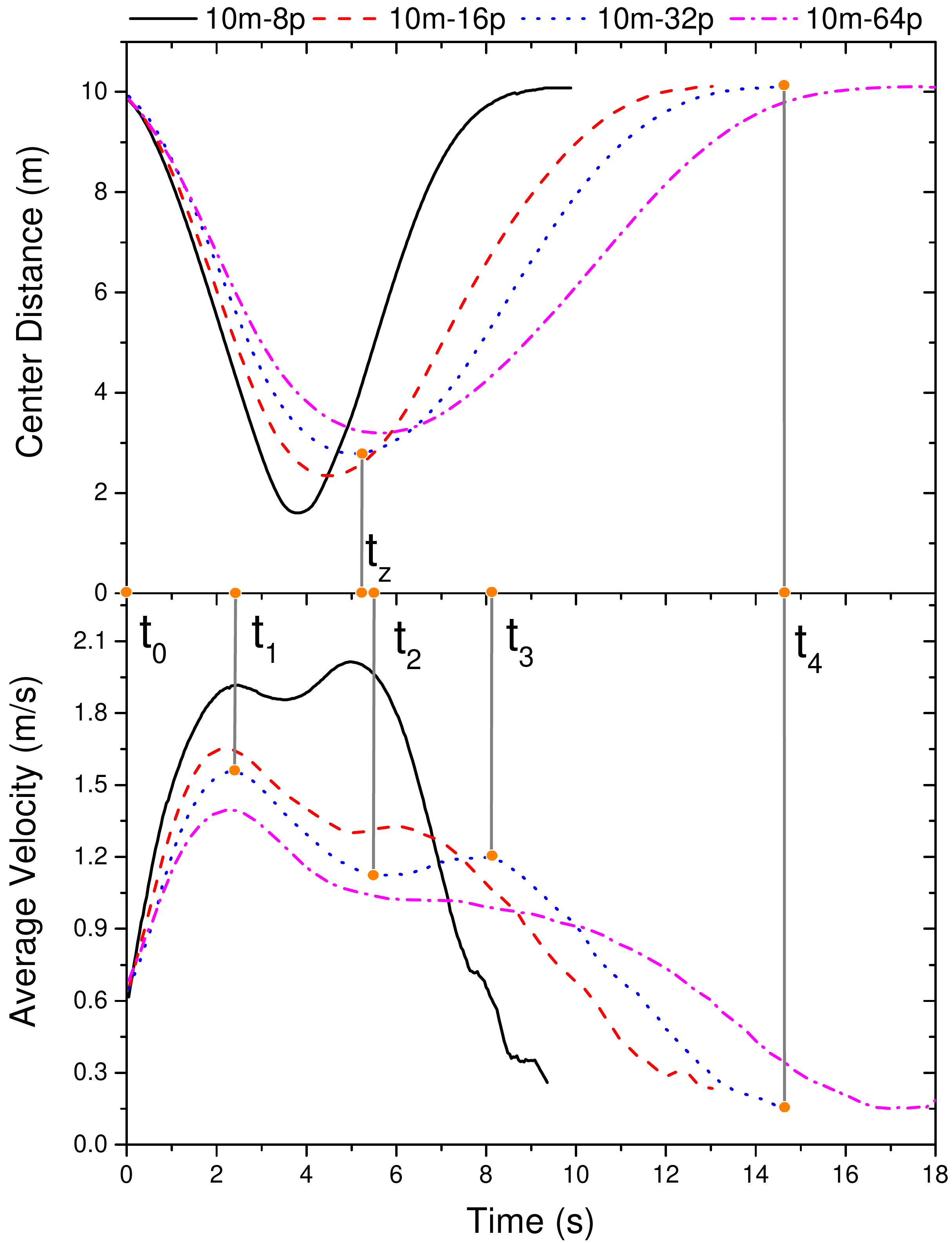}} 
\caption{Time series of center distance and speed.} \label{figtimeseries} 
\end{figure}

Fig. \ref{figtimeseries} shows the time series of average center distance and average velocity in the experiments. The time series of average center distance is a U-shaped curve overall, and the specific shape is formulated since the circle center is the midpoint of the shortest route of pedestrians. The minimal average center distance appears later and larger with the growing of pedestrian counts. The increasing count of pedestrians leads to a broader congestion region in the center area, so the average center distance is generally greater. Meanwhile, the speed of a pedestrian is likely to be affected by his/her surrounding, and the growing pedestrian counts result in a more crowded motion situation. To adapt the more crowded situation, pedestrians are likely to adjust the speed (mainly reduce the speed), and the minimal average center distance is thus less. 
% a little bit doubt

The basic pattern of the time series of average velocity usually contains two maximal values and one minimal value (Fig. \ref{figtimeseries}). Through exploring the time series of average velocity, five critical time points are defined and denoted in Fig. \ref{figtimeseries}, taking the 10m-32p experiment as an example. $t_0$ indicates the starting moment of the experiment. $t_1$ is the corresponding time of the first maximal value. $t_1$ is generated when the free acceleration process at the initial stage is interrupted. $t_2$ represents the corresponding time of the minimal velocity. It is noted that $t_2$ is fairly close to the time of the minimum value of the center distance. It makes sense since the moment of the minimal value of the center distance corresponds to the most crowded moment. $t_3$ denotes the corresponding time of the second maximal value. $t_3$ is generated when some pedestrians just accelerate from the most crowded situation while some other pedestrians have reached the destinations. $t_4$ is the ending moment of the experiment. In the case, the second average velocity increase might be counterbalanced and the second maximal value turns out to be not so obvious in some cases. Accordingly, the experiment can be divided into four stages. In the first stage($t_0$ to $t_1$), the pedestrians start off from the starting point and accelerate to the maximal velocity until being affected by other pedestrians. The second stage starts from $t_1$ and ends at $t_2$. The approaching of the center point makes the center region an even more crowded area, so the average velocity drops. The period from $t_2$ to $t_3$ is the third stage that the most serious conflicts and congestion have been managed and the average velocity makes a slight increase. In the last stage ($t_3$ to $t_4$), pedestrians reach their destinations one after another.

\section{Model evaluation}\label{section 4}

% 处理冲突和拥挤是行人动力学的核心问题 相应的参考文献
% Motivation of Evaluation, 我们实验的两个关键特点都比较适合于做一个评价体系。这一点还是要讲出来

The evaluation is a core step for exploring the model applicability, and an appropriate and quantitative evaluation framework is of great significance. The circle antipode experiment generally owns a serious conflicting and congested situation which can be quite challenging for pedestrian models, and the ability to deal with complicated situation shall be a most fundamental and critical index for the evaluation of pedestrian models. In addition, the symmetric experiment situation contributes to taking advantages of the trajectory data and providing more possibilities of quantitative evaluation. % 充分利用每个人的表现，可以观察行人的一般表现

\subsection{Evaluation framework}

\begin{figure}[!ht]
\centering
\includegraphics[width=0.7\textwidth]{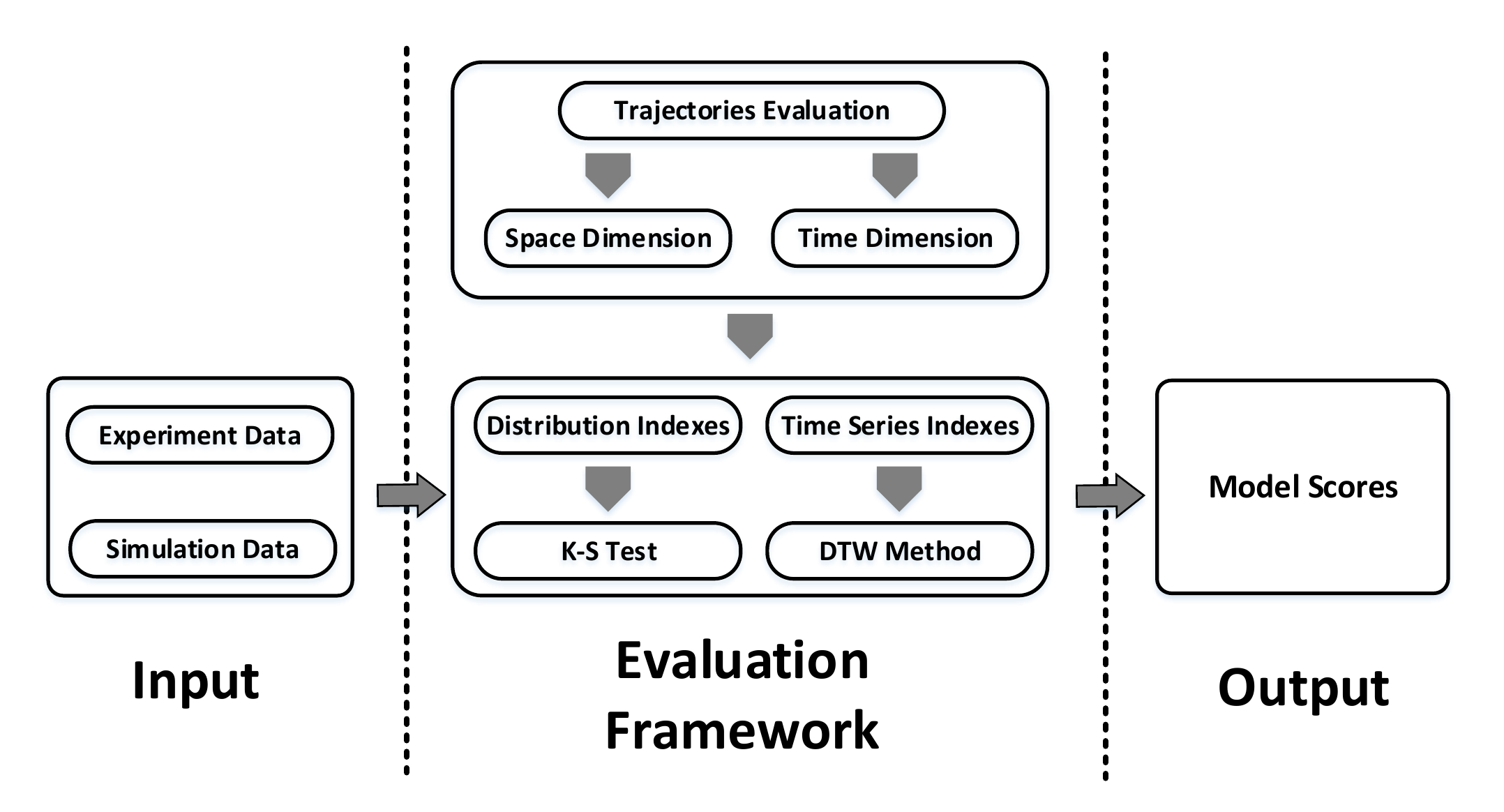}
\caption{The circle antipode experiment based evaluation procedures.}\label{figevaluation}
\end{figure}

% To take advantages of the challenging experiment and the symmetric experiment situation, a pedestrian model evaluation framework is formulated in the section. Several indexes in space and time dimensions are introduced for the evaluation of pedestrian simulation models. 
% 需要讲一讲我们这个评价方法的主要逻辑是什么，有什么关键的步骤和要素。

A pedestrian model evaluation framework is formulated in the section through the application of the challenging and symmetric circle antipode experiments (see Fig. \ref{figevaluation}). The framework contains six indexes in space and time dimensions, and two types of evaluation methods are adopted for analyzing the distribution indexes and the time series indexes, respectively. The model score can be calculated by introducing the experiment and the simulation results of the circle antipode experiment into the evaluation framework.

\subsubsection{Indexes}

The aim of a pedestrian model evaluation is to investigate to what extent the practical pedestrian behaviors can be reproduced by the simulation model. The symmetric and precise route trajectories extracted from the circle antipode experiments formulate the base data to reflect the motion properties. Investigating the spatial properties of pedestrian trajectories (Fig. \ref{figtrajectories} and Fig. \ref{figtrarotation}) can be the most effective evaluation method. Two measures including route length and route potential are used to estimate the spatial properties and differences of routes. 

\textbf{Distribution of route length.}
Usually, the longer route length iimplies a more radical detour route. The route length (see Eq. \ref{eq routelength}) can be regarded as an critical indicator for the spatial properties of route, and the distribution can be found in Fig. \ref{figdislength}.

\textbf{Distribution of route potential.}
The route potential (see Eq. \ref{eq potential}) is used to describe the detour features of pedestrian route, and the distribution of potential is shown in Fig. \ref{figdispotential}.

The combination of the route length and the route potential enhances the recognition of the features of a pedestrian route in the space dimension. However, the spatial features can't completely reflect the pedestrian behaviors in motion. Therefore, along with the two indicators, four additional measures are adopted.

\textbf{Distribution of travel time.}
The travel time is expressed in Eq. \ref{eq travel time}, and Fig. \ref{figdistime} illustrates the distribution.

\textbf{Distribution of speed.}
The speed is defined in Eq. \ref{eq speed}, and Fig. \ref{figdisvelocitysize} demonstrates the distribution.

\textbf{Time series of center distance.}
The time series of center distance denoted in Eq. \ref{eq average distance} (Fig. \ref{figtimeseries})can be applied to measure the degree of crowdedness during an experiment. 

\textbf{Time series of average velocity.}
The time series of average velocity introduced in Eq. \ref{eq average velocity}  (Fig. \ref{figtimeseries})is a measurement index of pedestrian status.

\subsubsection{Evaluation method} \label{section evaluationmethod}

To evaluate the similarity level between the experimental data and the simulation results, the Kolmogorov-Smirnov (K-S) test method and the dynamic time warping (DTW) distance method are introduced for the evaluation of the distribution data and the time series data, respectively. 

\begin{table}[!ht]
\centering
\caption{Evaluation Method I} \label{tab2}
\begin{tabular}{lp{15cm}} 
\toprule
\multicolumn{2}{c}{Method I (The K-S test evaluation method for experimental and simulation distribution data)} \\ 
\midrule
\textbf{Input:} 		&	Experimental distribution data set $\bm{Z}^{\rm{E}}$ and simulation distribution data set $\bm{Z}^{\rm{S}}$. \\
\textbf{Output: }	&	Normalized similarity evaluation score $S$. \\ 
\textbf{Step 1 }		&	Obtain the experimental cumulative distribution function $F^{\rm{E}}(x) $ and the simulation cumulative distribution function $F^{\rm{S}}(x) $ .\\
			&	$F^{\rm{E}}(x) = \frac {1} {n} \sum_{i=1}^n B(z_i^{\rm{E}},x),$ \quad $F^{\rm{S}}(x) = \frac {1} {n} \sum_{i=1}^n B(z_i^{\rm{S}},x), \quad z_i^{\rm{E}} \in{ \bm{Z}^{\rm{E}}}, z_i^{\rm{S}} \in{ \bm{Z}^{\rm{S}}}$ \\
			&	where $ B(z_i,x)$ equals to 1 if $z_i \leq x$ and equals to 0 otherwise. \\
\textbf{Step 2 }		&	Calculate the K-S statistic $D_{ks}$.\\
			&	$D_{ks} = \sup \limits_{x} |F^{\rm{E}}(x) - F^{\rm{S}}(x)|$, where $\sup \limits_{x}$ is the supremum of the set of distances. \\
\textbf{Step 3 }		&	Calculate the $p$ value of the K-S statistic and the normalized similarity evaluation score $S$.\\
			&	$ S =1/(1 - \log_{10}⁡ p )$. \\
\bottomrule 
\end{tabular}
\end{table}

\begin{table}[!ht]
\centering
\caption{Evaluation Method II} \label{tab3}
\begin{tabular}{lp{15cm}} 
\toprule
\multicolumn{2}{c}{Method II (The DTW distance evaluation method for the experimental and simulation time series data)} \\ 
\midrule
\textbf{Input:} 		&	Experimental time series data sequence set $\bm{Z}^{\rm{E}}$ and simulation time series data sequence set $\bm{Z}^{\rm{S}}$. \\
\textbf{Output: }	&	Normalized similarity evaluation score $S$. \\ 
\textbf{Step 1 }		&	Obtain the experimental time series sequence and the simulation time series sequence.\\
			&	Suppose that there are $a$ sets of time-series sequences exist in $\bm{Z}^{\rm{E}}$ and $b$ sets of time-series sequences exist in $\bm{Z}^{\rm{S}}$. $\bm{Z}_x^{\rm{E}}$ and $\bm{Z}_y^{\rm{S}}$ are two time sequences in the set $\bm{Z}^{\rm{E}}$and $\bm{Z}^{\rm{S}}$ and respectively own $m$ and $n$ symbols.\\
\textbf{Step 2 }		&	Calculate the DTW distance between the time series $\bm{Z}_x^{\rm{E}}$ and $\bm{Z}_y^{\rm{S}}$.\\
			&	Note that $\bm{z}_x^E(i)$ and $\bm{z}_y^S(j)$ denote two symbols in the set $\bm{Z}_x^{\rm{E}}$ and $\bm{Z}_y^{\rm{S}}$, respectively. $d\left(\bm{z}_x^E(i), \bm{z}_y^S(j)\right)$ is the Euclidean distance between the two symbols, i.e., $d\left(\bm{z}_x^E(i), \bm{z}_y^S(j)\right) =|\bm{z}_x^E(i)-\bm{z}_y^S(i)|$. \\
			&	1: \quad \textbf{for} \ $i$ = 1 to $m$\\
			&	2: \quad \quad\quad $DTW[i, 0] = \rm{infinity}$\\
			&	3: \quad \textbf{for} \ $i$ = 1 to $n$\\
			&	4: \quad \quad\quad $DTW[0, i] = \rm{infinity}$\\
			&	5: \quad $DTW[0,0] = 0$\\

			&	6: \quad \textbf{for} \ $i$ = 1 to $m$\\
			&	7: \quad \quad\quad \textbf{for} \ $j$ = 1 to $n$\\
			&	8: \quad \quad \quad\quad\quad $DTW[i, j] = d\left(\bm{z}_x^E(i), \bm{z}_y^S(j)\right)+ \min (DTW[i-1,j], DTW[i, j-1], DTW[i-1,j-1])$\\
			&	9: \quad \textbf{return} $DTW_{x, y} = DTW[m,n]$\\
\textbf{Step 3 }		&	Calculate the average DTW distance between the time series set $\bm{Z}^{\rm{E}}$ and $\bm{Z}^{\rm{S}}$. \\
			&	1: \quad \textbf{for} \ $x$ = 1 to $a$\\
			&	2: \quad \quad \quad \textbf{for} \ $y$ = $1$ to $b$\\
			&	3: \quad \quad \quad \quad \quad $DTW = DTW + DTW_{x,y}$\\
			&	4: \quad \textbf{return} $DTW = DTW / (a \cdot b ) $.\\
\textbf{Step 4 }		&	Calculate the normalized similarity evaluation score $S$. \\
			&	$ S =1/(1 + \log_{10}⁡ (1+DTW))$.\\
\bottomrule
\end{tabular}
\end{table}

The K-S test \citep{MasseyJr1951, Young1977} is a nonparametric test that is usually used to compare a sample with a reference probability distribution or to compare two samples, and it is sensitive to differences in both location and shape of the experimental cumulative distribution functions of the samples. Considering the specific condition of our evaluation, the K-S test can be used to quantify the differences between the simulation results and the experimental data. The specific procedures of the sample data evaluation are described in Table. \ref{tab2}. In general, the null hypothesis in the K-S test is given as that the experimental data and the simulation data come from the identical population, while the alternative hypothesis is that there are obvious differences between the populations of the two distribution sets. Considering the challenging circle antipode experiment, obvious differences are likely to be found between the practical simulation data and the experimental results. As a result, the practical p-value of the K-S test is likely to be quite small ($<0.05$) despite the range of p-value is 0 to 1. Considering the fact, a normalization process is introduced in step 3 of Table \ref{tab2} for the recognition and analysis of the results of K-S test. In the evaluation framework, the distributions of route length, route potential, travel time and speed are evaluated with the K-S test.

The DTW method \citep{PiyushShanker2007, Taylor2015} is a widely-used algorithm for the measure of similarity between two sequences (e.g., time series), and it has been applied to temporal sequences like audio, video and graphics data. Generally, the DTW method can calculate an optimal match between two sequences with certain restrictions. The sequences are "warped" non-linearly in the time dimension to determine a measure of the similarity independence of certain non-linear variations. The specific procedures for the DTW distance calculation are given in Table. \ref{tab3}. In the DTW distance method, the result varies from 0 to positive infinity, and a normalization process of the DTW value is also proposed to make a further analysis of the similarity between the experimental data and the simulation results. In the section, the time series of center distance and the time series of average velocity are evaluated with the DTW distance method.

During the formulation of the evaluation framework, the reliability of the evaluation indexes is another critical problem. Actually, the reliability mentioned here is mainly regarding the stability of the related indexes in repeated experiments. To investigate the reliability of the indexes, the verification processes are provided in \ref{appendix a}.

\subsection{Model and evaluation}

The social force model \citep{Helbing1995, Helbing2000} is widely-used in recent decades and famous for its simplicity, extendibility, and reproduction of the famous self-organized phenomena. In the section, a traditional social force model and a simple modified social force model are formulated to simulate the pedestrian motion in the circle antipode experiment, and their simulation results are analyzed with the evaluation framework. Through the evaluation of the two models, the performance of the evaluation framework can be obtained and investigated. 
% 需要介绍为什么引入这么两个模型，你这里没有讲清楚的。

\subsubsection{Traditional social force model} \label{section socialforcemodel}

In the social force model \citep{Helbing1995, Helbing2000}, the pedestrian is regarded as a kind of particle which is driven by several kinds of forces. The first kind of force reflects the desire of a pedestrian to reach the destination as quickly as possible. As a result, the pedestrian would like to move to the destination without detours and select the shortest route. In the case, the desired direction of pedestrian $P_i$ is 
\begin{equation} \label{eq original driving force direction}
\vec{e}_i^{des} = (\vec{l}_i^{des} - \vec{l}_i)/ \|\vec{l}_i^{des} - \vec{l}_i\|,
\end{equation}
where $\vec{l}_i^{des}$ indicates the destination position of pedestrian $P_i$, and $\vec{l}_i$ represents the actual position of pedestrian $P_i$. Each pedestrian owns a desired speed $v_i^0$, so the pedestrian would accelerate to a certain speed if not being interrupted by other pedestrians or obstacles. Due to the impact of other pedestrians and obstacles, the actual velocity usually varies from the desired velocity and the desire to maintain the desired velocity can be represented as, 
\begin{equation} \label{eq original driving force}
\vec{F}_i^{drv} = m_i (v_i^0 \vec{e}_i^{des} - \vec{v}_i)/\tau_i,
\end{equation}
where $\tau_i$ is a relaxation time for pedestrian $P_i$ to adjust the velocity, and it equals to 0.5 s in the work.

The second kind of force is the interaction force between pedestrians. Generally, a pedestrian prefers to keep a certain distance from other pedestrians around, and the repulsive force is thus generated. Note that the repulsive effect between pedestrians is not contradicted with some attraction interactions based on group behaviors or following behaviors, and the attraction force is not discussed in this paper. The repulsive force here is a more general interaction between unfamiliar pedestrians and increases with the approaching of them, and the repulsive effect between pedestrian $P_i$ and $P_j$ can be calculated as follows,
\begin{equation} \label{eq pedestrian interaction force}
\vec{F}_{ij}^{ped} = \left(A\exp{(-(r_{ij}-r_i-r_j)/B)}+k_n g(r_{ij}-r_i-r_j)\right)\vec{e}_{ij}^n + k_t g(r_{ij}-r_i-r_j) \vec{e}_{ij}^t,
\end{equation}
where $r_{ij}$ is the distance between pedestrian $P_i$ and $P_j$. $r_i$ and $r_j$ are the radius of pedestrian $P_i$ and $P_j$, respectively. The magnitude parameter $A$ and fall-off length $B$ are two constants that determine the strength and range of the social interaction, and the parameters take values of 2000 N and 0.08 m, respectively. $k_n$ and $k_t$ are the normal and tangential elastic restorative constants, and they respectively are set to be 120000 N/m and 240000 kg/m/s. $\vec{e}_{ij}^n$ refers to the unit vector from pedestrian $P_i$ to $P_j$, and $\vec{e}_{ij}^t$ indicates the perpendicular direction to $\vec{e}_{ij}^n$. $g(r_{ij}-r_i-r_j)$ equals to 1 when $r_{ij}-r_i-r_j \geq 0$ and equals to 0 otherwise.

The third kind of force stands for the interaction force between pedestrian and obstacle, and it is analogously formulated as, 
\begin{equation} \label{eq wall interaction force}
\vec{F}_{iw}^{obs} = (A\exp{(-(r_{iw}-r_i)/B)}+k_n g(r_{iw}-r_i))\vec{e}_{iw}^n + k_t g(r_{iw}-r_i) \vec{e}_{iw}^t,
\end{equation}
where $r_{iw}$ is the distance between pedestrian $P_i$ and the wall.

Based on the three kinds of forces, the motion of pedestrian $P_i$ is formulated as, 
\begin{equation} \label{eq result velocity}
d\vec{v}_{i}/dt = (\vec{F}_i^{des} + \sum_{M_i} \vec{F}_{ij}^{ped} + \sum_W \vec{F}_{iw}^{obs})/m_i,
\end{equation}
where $M_i$ and $W$ are two sets that contain the other pedestrians and the walls, respectively.

\subsubsection{Voronoi based social force model}

The primary motivation for the modification is to investigate the performance of the evaluation framework. Therefore, we propose a simple modification inspired by the Voronoi diagram method \citep{Xiao2016, Xiao2018, Qu2018} to improve the conflict management ability of the traditional social force model. The traditional driving force owns a special feature that its direction points to the destination all the time. The setting of driving force indeed reflects the desire for approaching the destination, but it also brings some disadvantages in reproducing the practical conflict and congestion avoidance behaviors. To deal with the potential problem, an alternative driving force direction is proposed based on the shape characteristics of the Voronoi diagram.

\begin{figure}[!ht]
\centering
\subfloat[ Voronoi diagram.] { \label{figvoronoi1} \includegraphics[width=0.3\textwidth]{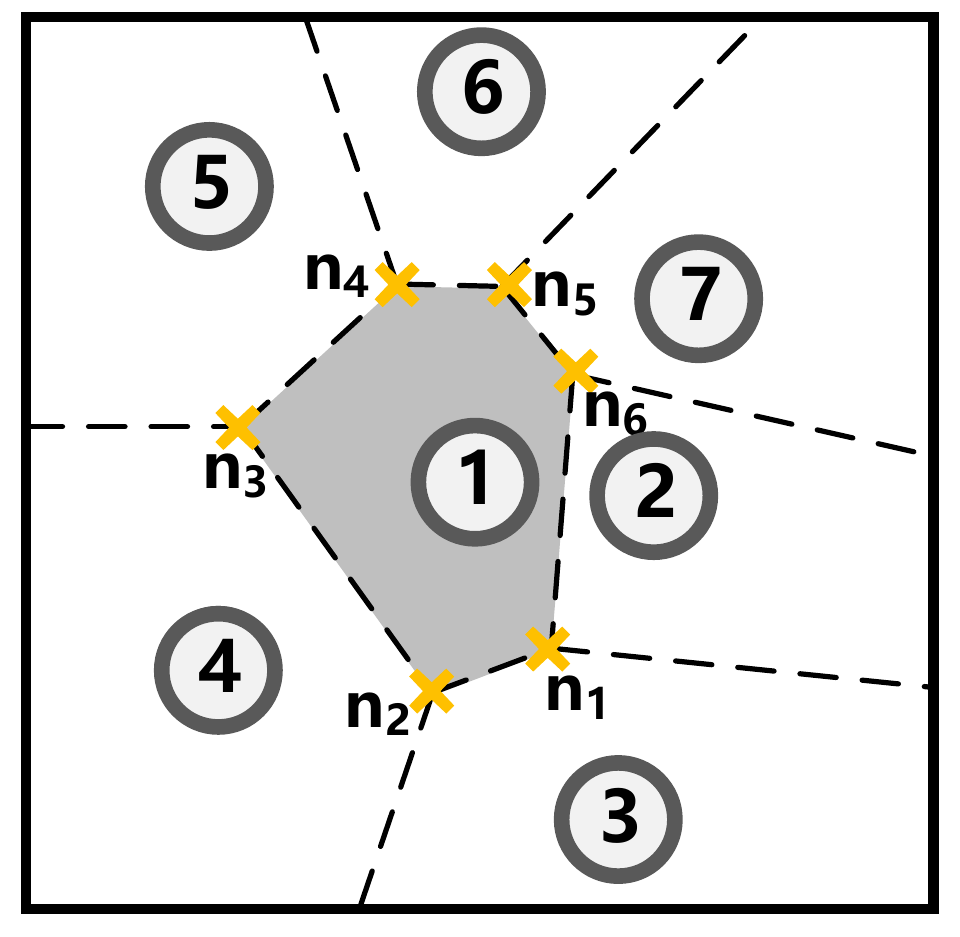}}
\subfloat[ Modified desired directions.] { \label{figvoronoi2} \includegraphics[width=0.3\textwidth]{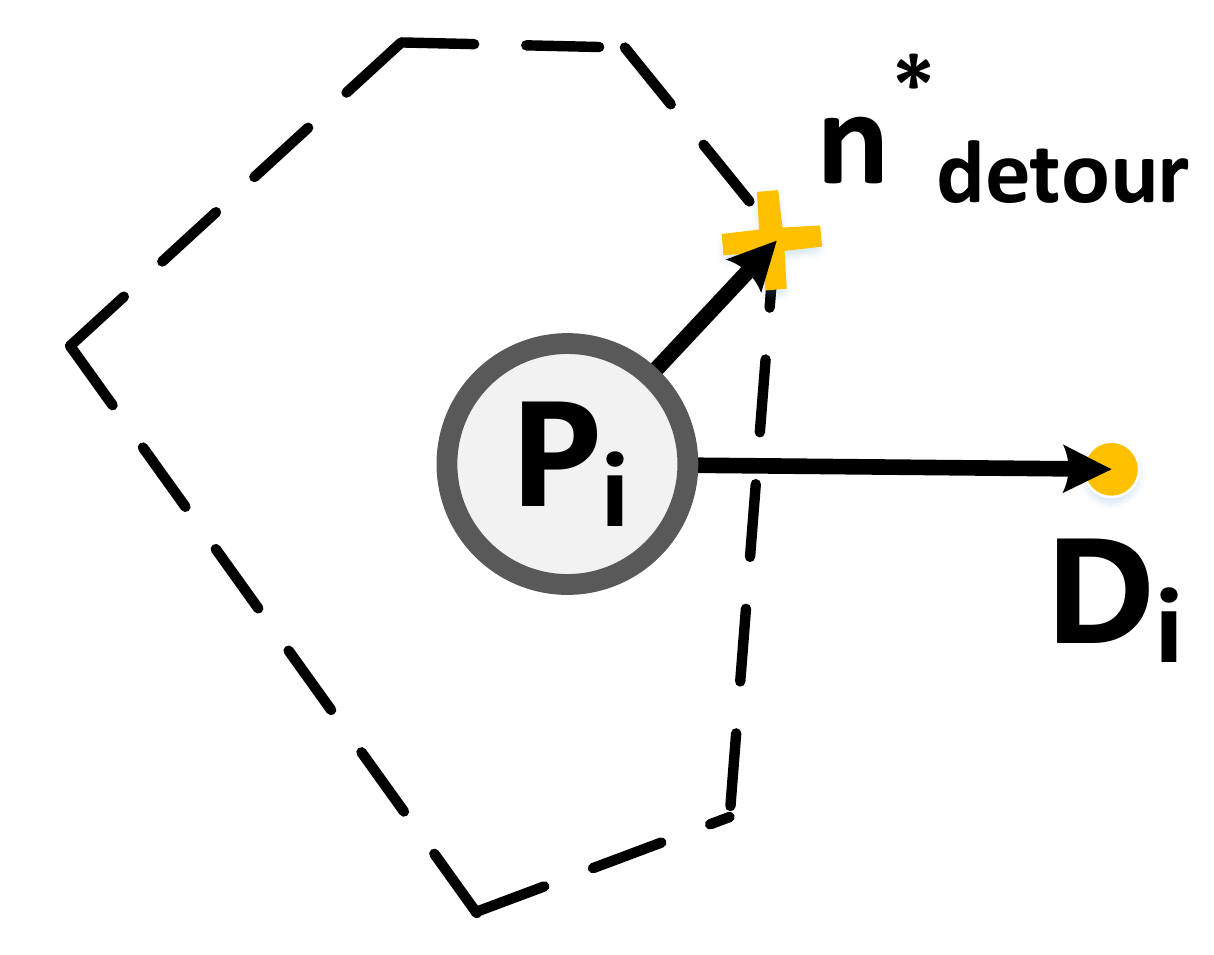}} 
\caption{Illustrations of Voronoi diagram and the modified desired direction. The gray circles represent the pedestrians. The Voronoi diagram are plotted with the dashed lines. The yellow forks represent the Voronoi nodes, and the yellow circle represents the destination of pedestrian.} \label{figVoronoi}
\end{figure}

The Voronoi diagram is a partitioning of space into regions based on distance to points in a specific subset, in other words, each region contains all the points closer to the related particle than to others. Based on the special geometry characteristics, the Voronoi diagram has been used in many areas including networking, biology and motion planning. Also, the Voronoi diagram shows potential in the investigation of pedestrian dynamics, and its shape characteristics could inspire possible direction choices for avoiding conflicts and congestions. As shown in Fig. \ref{figVoronoi}, 7 pedestrians represented by the circles are initialized in the square space, and the Voronoi diagram of the pedestrians is generated. The gray area is the Voronoi cell of pedestrian $P_1$, and it owns 6 Voronoi nodes. It is found that the direction to the Voronoi node indicates a potential direction to pass through the adjacent pedestrians. For instance, for pedestrian $P_1$, the direction to Voronoi node $n_1$ shall be a potential direction to pass through pedestrian $P_2$ and pedestrian $P_3$. In the same way, the 6 Voronoi nodes represent 6 possible detour directions.

Moreover, how to choose the optimal detour direction from the Voronoi nodes should be further explored. Here, several factors, e.g., destination, velocity, and deflection, are taken into consideration, and the optimal direction is determined by, 
\begin{equation} \label{eq optimal voronoi node}
n_i^{*} = \underset{n_{j}\in N_{i}}{\mathrm{\argmax}}(\vec{e}_i\cdot \vec{e}_{ij}/\rho^n_j),
\end{equation}
where $N_i$ represents the set of the Voronoi nodes of pedestrian $P_i$. $\vec{e}_i$ is the unit vector of velocity of pedestrian $P_i$, and $\vec{e}_{ij}$ is the unit vector from pedestrian $P_i$ to pedestrian $P_j$. $\rho^n_j$ is the local density of Voronoi node $n_j$. In this paper, the local density of a Voronoi node is defined as the average value of the densities of its related pedestrians. For example, the related pedestrians of Voronoi node $n_1$ include pedestrian $P_1$, $P_2$ and $P_3$. Moreover, the local density of a pedestrian is defined as the reciprocal of the area of its corresponding Voronoi cell. For instance, the local density of pedestrian $P_1$ equals to $1/a_1$, where $a_1$ indicates the area of the shadow region. In conclusion, the alternative desired direction is $\vec{e}^{\rm{dtr}}=\overrightarrow{P_in_i^*}/\|\overrightarrow{P_in_i^*}\|$.

With the introduction of an alternative desired direction which aims for the reasonable avoidance of conflicts and congestions, a further issue to answer is how to choose between the two types of desired directions. Actually, the problem is to recognize the critical condition between the choice of the original desired direction and the alternative one. In the case, a judgment formula is proposed,
\begin{equation} \label{eq critical condition}
C = d_{if} - \tau_i (\vec{v}_i - \vec{v}_f) \vec{e}_{if},
\end{equation}
where $\vec{v}_i$ and $\vec{v}_f$ are the velocities of pedestrian $P_i$ and its front pedestrian, respectively. $\tau_i$ is a relaxation time for pedestrian $p_i$. $d_{if}$ refers to the distance between them, and $e_{if}$ is the unit vector from pedestrian $P_i$ to its front pedestrian. Note that the front pedestrian is determined as the corresponding pedestrian in the Voronoi cell of the direction to the destination. In the modified social force model, the desired direction choice is modified,
\begin{equation} \label{eq optimal desired direction}
\vec{e}^{des}=\left\{
\begin{array}{rcl}
\vec{e}^{des}, & & {C \geq 0 }\\
\vec{e}^{dtr}, & & {\rm{otherwise} }\\
\end{array}.\right.
\end{equation}

\subsubsection{Simulation and evaluation}

In traditional simulation scenes (e.g. corridor and bottleneck), both models have solid performance \cite{Helbing1995,Helbing2000,Xiao2016,Qu2018}. The simulated fundamental diagram agrees well with the experimental results, and many typical phenomena (e.g., lane formation and arching phenomenon) are reproduced in both models (see \url{http://pedynamic.com/voronoi-based-social-force-model/}). Especially, the Voronoi based social force model works well in handling the conflicts with other pedestrians.

\begin{figure}[!ht]
\centering
\includegraphics[width=0.8\textwidth]{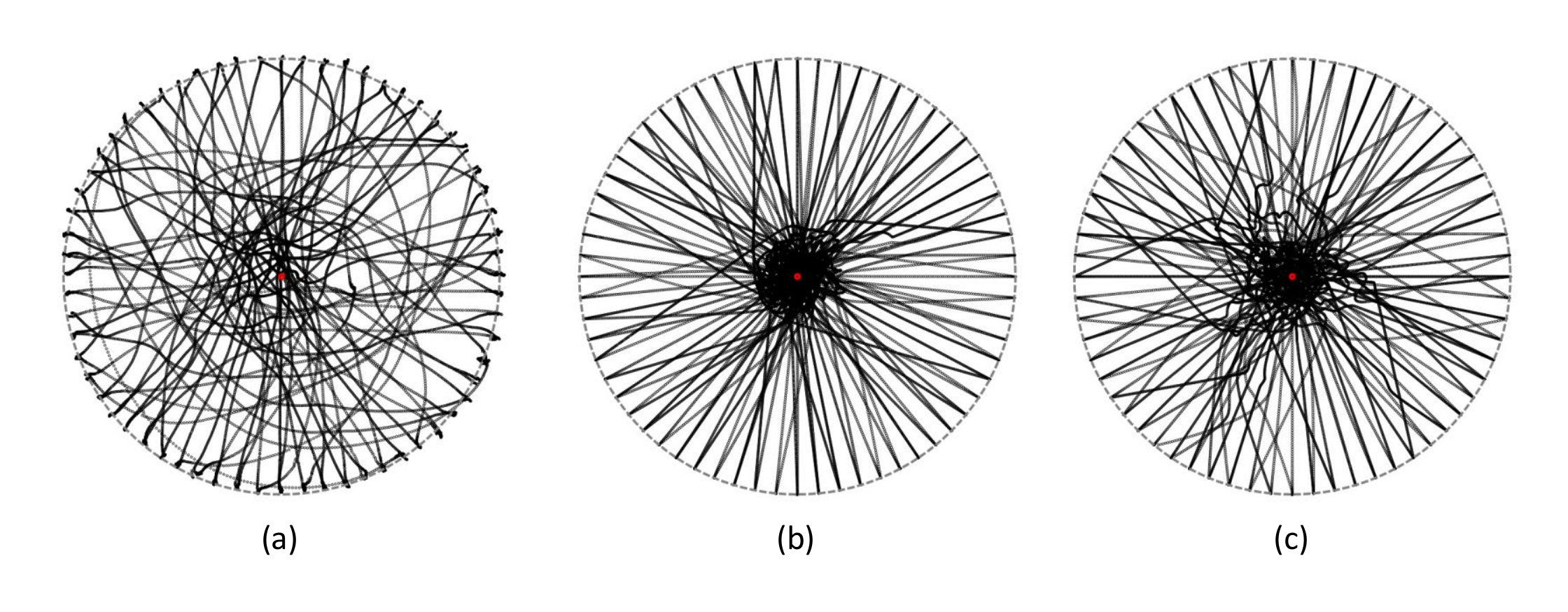}
\caption{Illustrations of pedestrian trajectories. (a) Trajectories in the empirical experiments. (b) Trajectories in the traditional social force model. (c) Trajectories in the modified model.}\label{figSimulation}
\end{figure}

In this work, the 10m-64p experiment is chosen as the corresponding evaluation experiment scenario, and it has been repeated 100 times in the simulations for the traditional social force model and the modified social force model, respectively. As mentioned in the model section, the improved model outperforms the traditional one by only considering a rather simple but critical factor, an alternative desired direction. The characteristics of the alternative desired direction can be particularly useful in the circle antipode experiments where serious conflicts occur. The pedestrian trajectories in the simulations are extracted with 25 frames per second and drawn in Fig. \ref{figSimulation}. From a qualitative perspective, more frequent and intense detour behaviors are found in the improved model. Apparently, the modification of the desired direction is beneficial to figure out conflicts and reproduce reasonable pedestrian behaviors. Besides, more videos about the simulations are shown on our website, \url{http://pedynamic.com/circle-antipode-simulation/}.

\begin{figure}[!ht]
\centering
\includegraphics[width=0.6\textwidth]{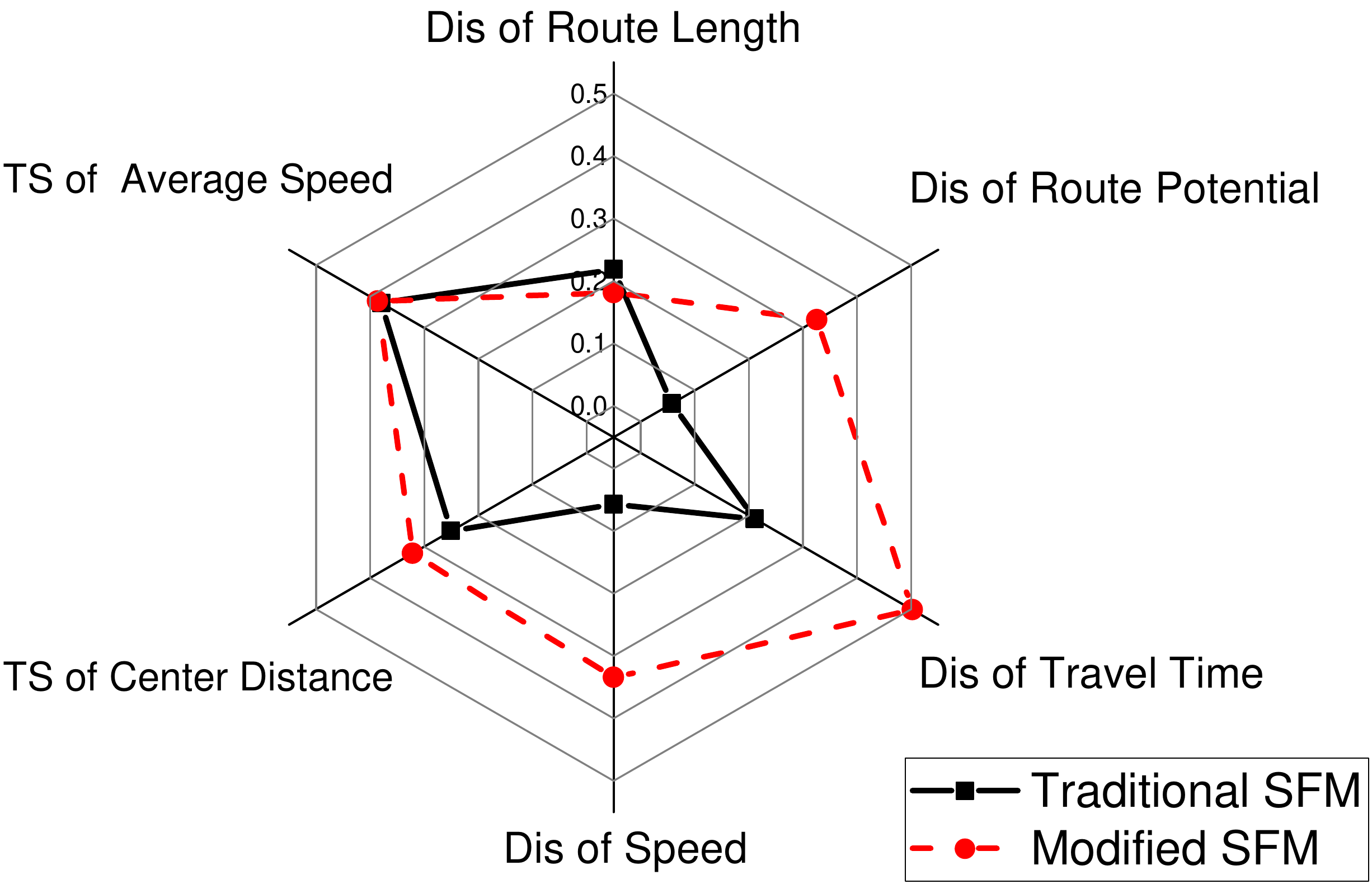}
\caption{Radar figure of the evaluation results. Note that Dis and TS represent the distribution indexes and time series indexes, respectively.} \label{figradar2}
\end{figure}

The related indexes are calculated based on the simulated pedestrian trajectories, and the evaluation results of the two models are presented in Fig. \ref{figradar2}. Apart from route length, the modified social force model gets higher ratings in the other indexes. The results make sense since more realistic behaviors especially the conflict avoidance behaviors can be reflected by the modified model. For the route length index, the mean values of both models are generally less than the experimental results due to the lack of the global route navigations. The pedestrians in the traditional social force model spend more time to deal with the conflicts and congestions in the center region, which increases the emerge probability of longer routes. Hence, the score of the route length index in the traditional social force model is greater. Generally, although the model modification is quite limited, it is still evident that the overall performance of the modified social force model is greater than that of the traditional one. That's to say, the improvements of the simple modification are reflected by the current evaluation framework.

\section{Discussion and prospect}\label{section 5}
% Detouring route, Direct Route 

In this paper, up to 32 sets of circle antipode experiments were performed to reveal the route navigation and conflict avoidance mechanisms of pedestrians. The specific pedestrian trajectories during the experiments were precisely recognized, and these trajectories are regarded as the main experimental data source in this work. The frequent and serious conflicts formulated in the center region are beneficial to the investigation of route navigation and conflict avoidance, and the symmetric experimental conditions in the experiments further enable more available quantitative analyses on pedestrian behaviors. Considering these features, the spatial distribution and several indexes based on the original and rotated trajectories are investigated. Several noticeable conclusions are found and further discussed as follows.

The spatial statistics of the rotated trajectories showed an apparent walking preference of pedestrians on the right side in the experiments. The side preference is considered to be related to the culture and regions \citep{Helbing2005,Moussaid2009}, and the practical habits in China confirm that our findings make sense. Similar to the side preference and the fundamental diagram \citep{Chattaraj2009}, the heterogeneous nature of pedestrians might lead to diverse pedestrian performance in the experiments. Therefore, considering more factors including ages, heights, nations, cultures and even psychological states etc, are useful for more reliable and convincing results in the circle antipode experiments. The symmetric experiment condition of the circle antipode experiment benefits the formulation of a public and open source database.

Another topic is regarding the distribution of the indexes. Current fitting results show that the route length distribution follows the log-normal distribution, the route potential distribution fits the exponential distribution, and travel time distribution and the speed distribution agree well with the normal distribution. The formulation mechanisms of these distributions are essentially different. For instance, a log-normal process is the statistical realization of the multiplicative product of many independent random variables, whereas the normal distribution is obtained by the sum of many independent processes. However, the formulation mechanisms for specific distribution of route length, route potential, travel time and speed are not so clear and need further discussions.

% In this section, we discuss two kinds of approaches: one where information is obtained from a central authority, and another where information is obtained by communicating with other agents.
There are basically two types of route choices in the circle antipode experiment: one passing through the center region, and the other detouring around the center region. It is found that the pedestrians taking the shortest routes experience long average travel times, while the detouring pedestrians even arrive at their destination faster. The crowding effect is believed to be the core reason for the non-positive correlation between route length and travel time. The attractive short routes are usually crowded with pedestrians, thus they are intractable and time-consuming to pass through; while the detour routes seem to be less popular, where pedestrians avoid the crowdedness and walk more smoothly. Besides, the travel time for each pedestrian shall be approximately the same according to the UE theory in the transportation distribution, which is not consistent with our results. Two kinds of causes might lead to the results. First, the experimental process especially the effect of congestion in the center region was not fully recognized by pedestrians, so over-many pedestrians chose the route passing through the center region. Second, the practical cost function of the route choice may contain more factors, e.g., energy saving, other than pedestrian travel time. Further experiments can be performed to investigate the specific reason for the phenomenon. For instance, the circle antipode experiments are substantially repeated with a fixed group of pedestrians. Through an analysis of the index changes over time, the reason for the length-time results can be further explored.

%在实验的分析部分 能做的东西 其实还有好多好多。。。 
% 文章的后半部分主要是关于模型评价
% 还有一个问题是 我们依据此实验 评价的模型 确定的参数体系 究竟能否适应其他场景呢？ 
% 没有提 我们的评价体系的 优势和特点 

What's more, an evaluation framework of the pedestrian model is formulated in the work in which experimental and simulation trajectories are the base data. In the framework, several space and time dimensional indexes, i.e., distribution of route length, distribution route potential, distribution of travel time, distribution of speed, time series of center distance and time series of average velocity, are introduced as indicators. The distribution indexes and time series indexes are respectively evaluated with K-S test and DTW method. This paper further applied the method to evaluate a traditional social force model and a simple modification. It is found that the basic performances of both models are reflected, and the simple modification improves the performance according to our evaluation framework. Since the realistic conditions of pedestrian dynamics are varied and complex, it has to be admitted that the effectiveness of the circle antipode experiments based framework would be quite limited. Still, the ability of route navigation and conflict avoidance is a core and maybe a most challenging problem in crowd dynamics, and the proposed evaluation framework based on the circle antipode experiments would be significant for the pedestrian model evaluation.

Indeed, significant differences can be found between the empirical and simulation trajectories, and a lot of room for improvement exists in the current pedestrian models. Through investigation of the simulation processes, the point is that a real pedestrian normally predicts the short route or the detour route from the very beginning, whereas it is to our knowledge very hard to achieve in current models due to the lack of global and dynamic routes. In sum, it also provides a potential research direction, about how to predict the conflicts and decide the route choice in a complex and dynamic situation for pedestrians and even robots.

\section*{Acknowledgment}\label{section acknowledgment}

The study is supported by the Foundation for Innovative Research Groups of the National Natural Science Foundation of China (Grant No. 71621001), National Key R\&D Program of China (Grant No. 2017YFC0803300), the National Natural Science Foundation of China (Grant No. 71771021, 71601017), State Key Laboratory of Rail Traffic Control and Safety (No. RCS2018ZQ001). 

\begin{appendix}

\section{Indexes verification}\label{appendix a}
\setcounter{table}{0}
\setcounter{figure}{0}

\begin{figure}[!ht]
\centering\includegraphics[width=\textwidth]{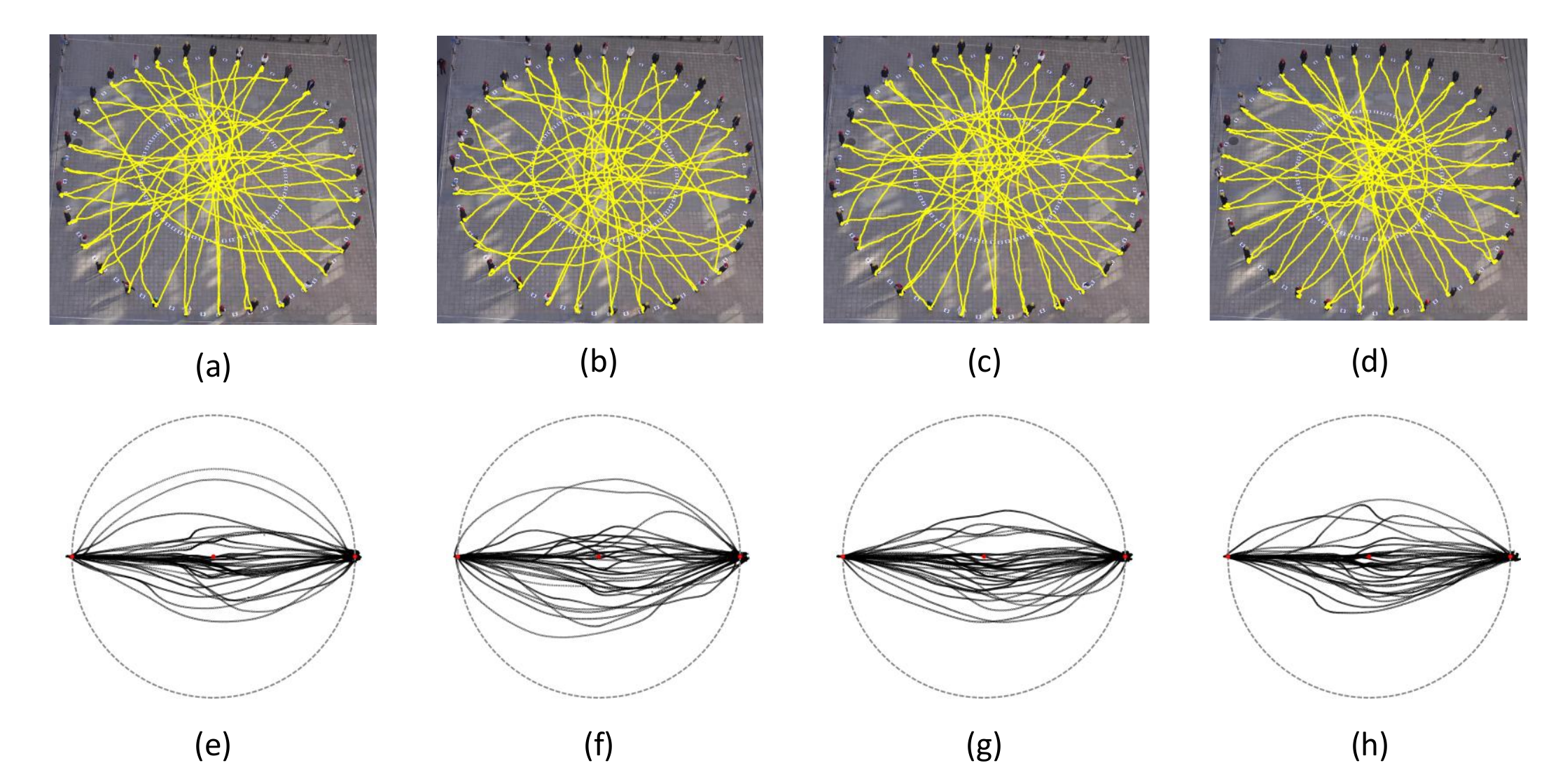}
\caption{The original and rotated trajectories of the repeated experiments in the 10m-32p experiments. (a) - (d) show the original trajectories of the repeated experiments. (e) - (h) show the rotated trajectories of the repeated experiments.} \label{figspace1}
\end{figure}

Fig. \ref{figspace1} shows the original ground trajectories and its responding rotated trajectories in the four repeated experiments of the 10m-32p experiments. In general, noticeable differences of the shape features of the pedestrian trajectories (both original and rotated) in the repeated experiments emerge, whereas the differences are significantly less than different types of experiments (Fig. \ref{figtrajectories} and Fig. \ref{figtrarotation}).

To test the stability of indexes in the repeated experiments, the Kruskal-Wallis (K-W) Test and the dynamic time warping (DTW) method are introduced. 

The Kruskal-Wallis test \citep{Kruskal1952, Breslow1970} is a non-parametric method to test whether samples originate from the same population, and it is appropriate to compare two or more samples with the same or different size. Note that a normal distribution of the residuals is not required for the K-W test since it is a non-parametric method. Generally, the null hypothesis assumes that the samples (or groups) are from identical populations, and the alternative hypothesis is that at least one of the samples (or groups) comes from a different population. The specific procedures of the method are given in Table. \ref{tab4}. The calculated $p$ value is shown in Table. \ref{tab6}. It is found that all the $p$ values are greater than 0.05, which means that the null hypothesis can not be rejected and the samples might come from identical populations. As a conclusion, the values of the four indexes are stable among the repeated experiments.

According to the features of the DTW method introduced in section \ref{section evaluationmethod}, it can also be introduced into the stability verification of the indexes in the repeated experiments. The specific procedures for stability investigation of the indexes are given in Table. \ref{tab5}. The results of the DTW method show that the two time series indexes are adequately stable among the repeated experiments.

\begin{table}[!ht]
\centering
\caption{Index stability investigation method I} \label{tab4}
\begin{tabular}{lp{15cm}} 
\toprule
\multicolumn{2}{c}{Method I (The K-W test method for the repeated experiments)} \\ 
\midrule
\textbf{Input:} 		&	Experimental distribution data set $\bm{Z}_{K}^{\rm{E}}$ from the $K$ repeated experiments.\\
\textbf{Output: }	&	$p$ value of the K-W test.\\ 
\textbf{Step 1 }		&	Obtain the rank of all data from repeated experiments.\\
			&	Here, $\delta_i(j)$ denotes the overall rank of $j$th data in the $i$th repeated experiment. \\
\textbf{Step 2}		&	Calculate the K-W statistics. \\
			&	$KW = (N-1) \sum_{i=1}^{K} n_i(\overline{\delta}_i - \overline{\delta})^2 / \sum_{i=1}^{K} \sum_{j=1}^{n_i} (\delta_{i}(j) - \overline{\delta})^2.$\\
			&	Where $\overline{\delta}_i$ denotes the average overall rank of the data in the $i$th repeated experiment and $\overline{\delta}_i = \sum_{j=1}^{n_i} (\delta_{i}(j)) $. $\overline{\delta} $ represents the average of all the $\delta_{i}(j)$ and $\overline{\delta} = 0.5(N+1)$. $n_i$ is the number of data in the $i$th repeated experiment. $N$ denotes the total number of data among repeated experiments. $K$ stands for the number of the repeated experiments\\
\textbf{Step 3}		&	Calculate the corresponding p-value of the K-W statistics.\\ 
			&	$p \approx Pr(\chi_{K-1}^2 \geq KW) $ \\
\bottomrule
\end{tabular}
\end{table}

\begin{table}[!ht]
\centering
\caption{Index stability investigation method II} \label{tab5}
\begin{tabular}{lp{15cm}} 
\toprule
\multicolumn{2}{c}{Method II (The DTW stability investigation method for the repeated experiments)} \\ 
\midrule
\textbf{Input:} 		&	Experimental time series data sequence set from the repeated experiments $\bm{Z}^{\rm{E}}$. \\
\textbf{Output: }	&	Average DTW distance $DTW$. \\ 
\textbf{Step 1 }		&	Obtain the time series sequences of the repeated experiments.\\
			&	Suppose that there are $a$ sets of time-series sequences in $\bm{Z}^{\rm{E}}$. $\bm{Z}_x^{\rm{E}}$ represents a time sequence in the set $\bm{Z}^{\rm{E}}$ and owns $m$ symbols.\\
\textbf{Step 2 }		&	Calculate the DTW distance between the repeated experiments.\\
			&	Note that $\bm{z}_x^E(i)$ and $\bm{z}_y^E(j)$ denote two symbols in the set $\bm{Z}_x^{\rm{E}}$ and $\bm{Z}_y^{\rm{E}}$ from the repeated experiments, respectively. $d\left(\bm{z}_x^E(i), \bm{z}_y^E(j)\right)$ is the Euclidean distance between the two symbols, i.e., $d\left(\bm{z}_x^E(i), \bm{z}_y^E(j)\right) =|\bm{z}_x^E(i)-\bm{z}_y^E(i)|$. \\
			&	1: \quad \textbf{for} \ $i$ = 1 to $m$\\
			&	2: \quad \quad\quad $DTW[i, 0] = \rm{infinity}$\\
			&	3: \quad \textbf{for} \ $i$ = 1 to $n$\\
			&	4: \quad \quad\quad $DTW[0, i] = \rm{infinity}$\\
			&	5: \quad $DTW[0,0] = 0$\\
			&	6: \quad \textbf{for} \ $i$ = 1 to $m$\\
			&	7: \quad \quad\quad \textbf{for} \ $j$ = 1 to $n$\\
			&	8: \quad \quad \quad\quad\quad $DTW[i, j] = d\left(\bm{z}_x^E(i), \bm{z}_y^S(j)\right) + \min (DTW[i-1,j], DTW[i, j-1], DTW[i-1,j-1])$\\
			&	9: \quad \textbf{return} $DTW_{x, y} = DTW[m,n]$\\
\textbf{Step 3 }		&	Calculate the average DTW distance between the repeated experiments. \\
			&	1: \quad \textbf{for} \ $x$ = 1 to $a-1$\\
			&	2: \quad \quad \quad \textbf{for} \ $y$ = $x+1$ to $a$\\
			&	3: \quad \quad \quad \quad \quad $DTW = DTW + DTW_{x,y}$\\
			&	4: \quad \textbf{return} \ $ DTW = 2 \cdot DTW / (a^2 -a ) $.\\
\bottomrule
\end{tabular}
\end{table}

\begin{table}[!ht]
\centering
\caption{Index analysis} \label{tab6}
\begin{tabular}{ccc}
\toprule
Index & Method & Value \\ 
\midrule
Route Length 				& K-W Test 		& 0.515 \\
Route Potential 			& K-W Test 		& 0.666 \\
Travel Time 				& K-W Test 		& 0.602 \\
Velocity 				& K-W Test 		& 0.577 \\
Time Series of Average Velocity 	& DTW Method 	& 17.400 \\
Time Series of Center Distance 		& DTW Method 	& 17.360 \\
\bottomrule
\end{tabular}
\end{table}
\newpage 

\end{appendix}

\section*{References}

\bibliography{main}
% \bibliographystyle{IEEEtran} %这是你要使用的格式,比如要投IEEE,就写IEEEtran
% \bibliography{IEEEabrv, circlegame}

\end{document}